\documentclass[journal]{IEEEtran}
\ifCLASSINFOpdf
\else
\fi
\usepackage{cite}
\usepackage{amsmath,amssymb,amsfonts}
\usepackage{algorithmic}
\usepackage{graphicx}
\usepackage{textcomp}
\usepackage{amsmath}
\usepackage{hyperref}
\usepackage{booktabs}
\usepackage{array}
\usepackage{alltt}
\DeclareUnicodeCharacter{2212}{-}
\usepackage[numbers,,sort&compress]{natbib}
\usepackage{subcaption}
\usepackage{parskip}

\begin{document}
%
\title{Public Transit of the Future: Enhancing Well-Being through Designing Human-centered Public Transportation Spaces}
%
%
%

\author{Yasaman Hakiminejad, Elizabeth Pantesco, and Arash Tavakoli
\thanks{Email: arash.tavakoli@villanova.edu}
}
%
%

\markboth{Journal of \LaTeX\ Class Files,~Vol.~14, No.~8, August~2015}%
{Shell \MakeLowercase{\textit{et al.}}: Bare Demo of IEEEtran.cls for IEEE Journals}
%



\maketitle

\begin{abstract}

Studies show that psychological effects are among one of the top concerns for public transportation users. While many Americans spend a significant portion of their time in public transportation spaces (7 billion trips in 2023), the impact of the design and maintenance of these spaces on user well-being has not been fully studied. In this study, we conducted a survey to better understand the effect of implementing different functional features on people's well-being and perceptual metrics (N=304). Participants were presented with six images depicting different cabin configurations, including (1) the current version of the cabin space from a regional train in the United States as a baseline, (2) the current version with added trash and visible wear and tear to illustrate a lack of maintenance, (3) an aesthetically enhanced version of the current cabin space with changes in material, seat arrangement, and presentation, (4) a bike rack-enabled version to accommodate multimodal transportation needs, (5) a version with an added workspace to meet personal and work-related needs, and (6) an improved version with added plants to demonstrate biophilic design. After viewing each image, participants well-being metrics (i.e., stress, valence, arousal, and creativity) and their public transportation perception metrics (i.e., perceptions of safety, reasonable cost, punctuality, comfort, passage of time, cleanliness, workspace utilization, and likelihood of recommending the service) were evaluated through standardized questionnaires. We used linear mixed-effect models (LMM) for analyzing the difference in the dependent variables (i.e., well-being and perceptual metrics) across the various design conditions as independent variables. Our results indicated that adding functional amenities (e.g., bike racks, workstations) and biophilic design elements (e.g., plants) leads to an overall enhancement in well-being and perceptual metrics. Conversely, low maintenance and upkeep worsened all measured well-being and perceptual metrics, underscoring the importance of good maintenance for user satisfaction. This research lays the ground for developing human-centered public transportation spaces that can lead to an increase in public transportation adoption and improved user satisfaction. 

\end{abstract}


%
\IEEEpeerreviewmaketitle

\section{Introduction}
\label{sec:introduction}
In today's urban lifestyle, a large portion of people spend their daily routines either in stations or within the cabins of public transportation. Whether they are commuting to work, connecting with loved ones, or exploring the city, these transportation spaces serve as temporary cabins for a significant portion of the population daily. In 2023 alone, Americans made more than 7 billion trips using public transportation \cite{noauthor_undated-tx}. Despite the vital role that these transportation infrastructures play in our daily lives and the significant portion of time and exposure various groups of people have to these spaces, there has been limited exploration into the impact of the design of public transportation spaces on riders' well-being, perceptions of comfort, safety, and cognitive related attributes.

From a safety perspective, using public transportation can be up to 10 times safer per mile compared to driving personal automobiles \cite{Public_Transportation_2016}. Despite this significant safety advantage, there is a widespread preference for cars over public transportation in the United States \cite{STEG200327}. While infrastructure shortcomings are often cited as a reason for this preference \cite{Transportation_Research_Board2001}, a considerable portion of the issue resides at the intersection of psychology and engineering \cite{Galdames2011-nj,Liao2017-gx}: Are these transportation spaces adequately designed to meet the needs of their users?

Numerous studies have delved into the needs and priorities of daily users of public transportation, particularly in the United States, and have examined why public transportation is not as widely used there as it is in Europe \cite{Transportation_Research_Board2001,doi:10.1080/01441649508716906}. These studies often cite factors such as the lower density of settlements, longer trip distances, and the lower price of motor fuel in the United States compared to Western Europe as reasons for the differences in adoption. However, there is a noticeable void in the literature regarding the effects of different cabin space designs and overall maintenance on passengers' experiences and well-being. Addressing this gap is vital, as the design and upkeep of public transportation spaces can significantly impact user satisfaction and overall usage rates, similar to any other built environment such as buildings, and urban spaces \cite{douglas2022physical}. Understanding how these factors influence passenger experiences can lead to improved public transportation systems that better meet the needs of users, thereby enhancing their overall well-being and encouraging greater use of public transit.

This paper is geared toward addressing this gap by taking the first steps in systematically quantifying the effect of various alternative designs of the same public transportation space on an array of metrics regarding user's well-being and space perception. More specifically, we conducted an online survey where participants were shown six different images of cabin spaces and were then asked to respond to a series of standard scales focused on different aspects of well-being (i.e., stress, emotion, and creativity), and perceptual metrics (i.e., the likelihood of recommending the cabin to others, the passage of time, desirability to work in, safety, comfort level, cleanliness, punctuality of service, and the reasonable cost associated with it). The images depicted a range of cabin configurations including (1) \textit{the current version} of the cabin space as a baseline taken from a regional train in the United States, (2) \textit{the current version with added trash and visible wear and tear} as an example of lack of maintenance, (3) \textit{an aesthetically enhanced version} of the current cabin space which involved changing the material, seat arrangement, and presentation, (4) a \textit{bike rack enabled} version featuring a bike rack as an example for a design that accommodates multimodal transportation needs, (5) another with an \textit{added workspace} as an example that accommodates personal and work-related needs, and lastly, (6) \textit{an improved version with added plants} as an example of having biophilic designs geared towards enhancing participants' well-being. By examining these varied designs, our objective is to gain a deeper understanding of how specific elements within the cabin environment influence passengers' perceptions and overall well-being. This comprehensive approach allows us to identify key factors that can enhance the user experience and inform future design improvements in public transportation spaces.

This study aims to address two main questions:
\begin{enumerate}
\item How do the design, features, and maintenance of public transportation spaces impact self-reported well-being metrics?
\item How do the design, features, and maintenance of public transportation spaces influence self-reported perceptions of safety, punctuality of service, desirability, comfort, and the perception of the passage of time?
\end{enumerate}

By exploring these questions, our goal is to uncover insights that can guide the design and enhancement of public transportation environments, ultimately improving user experience and encouraging increased use of public transport. This research aims to contribute to a deeper understanding of the factors that impact public transportation satisfaction and to offer practical recommendations for policymakers and designers seeking to develop more user-centered and appealing public transit systems.

In answering the aforementioned questions, we hypothesize that public transportation spaces significantly influence the well-being of their users as well as their perception of public transportation in the following ways:
\begin{enumerate}
    \item The design of the cabin can enhance people's well-being, including decreasing their perceived stress level and increasing their positive emotions and creativity. 
    \item The design of the cabin can enhance people's space perceptual metrics such as the feeling of safety, and cleanliness.
    \item Visual enhancements and the inclusion of functional features in the cabin can boost people's willingness to use public transportation and recommending it to others.
    \item Public transit systems that are equipped with cabins that undergo proper maintenance can positively affect people's well-being and increase their willingness to use public transportation.
\end{enumerate}

\section{Background}
\subsection{Well-being}
Well-being is a multifaceted construct that encompasses psychological, social, and physical dimensions of health and functioning. Given its associations with productivity, longevity, and overall quality of life, well-being is increasingly measured in studies aiming to inform public health interventions, improve policy, and guide infrastructure development (\cite{diener2018advances}). An important component of well-being is hedonic, or experiential, well-being, which involves the balance of positive and negative emotional states and is best measured by momentary assessments at a given point in time (\cite{NCCIH_2018,Kahneman1999-KAHWTF}). One approach to the assessment of emotional states is to use a dimensional framework, in which emotions are posited to occur along the two distinct axes of valence and arousal (\cite{barrett1998discrete,russell1999core}). Emotional valence refers to the hedonic tone of the emotion, from positive to negative, whereas arousal refers to the degree of activation or alertness associated with the emotion (\cite{barrett1998discrete}).

While not typically included in most formal definitions of well-being, several closely related constructs, such as stress and creativity, may offer additional insight into an individual's psychological state \cite{Fredrickson2001}. For instance, reports of psychological stress, or experiences that are appraised as "taxing or exceeding [one's] resources and endangering [one's] well-being” \cite{lazarus1984stress}, generally map on to the emotional dimensions of negative valence and high arousal. With regard to creativity, the broaden-and-build theory posits that an individual’s momentary thought-action repertoire, or the range of accessible responses to a given situation, expands when experiencing positive emotions (\cite{fredrickson2001role}). Levels of creativity would therefore be expected to increase with more positively-valenced emotions. Assessing psychological factors such as stress and creativity in conjunction with emotional valence and arousal may offer a more comprehensive picture of emotional well-being and, as such, are included as outcomes in the current study.

\subsection{Previous Works on the Effect of Public Transportation Spaces Design and Maintenance on User's Well-being and Perceptual Metrics}

Prior work has explored the potential benefits of enhancing public transportation spaces for public health and urban livability. For instance, Brown et al. showed that improving public transportation accessibility can increase physical activity, reduce stress, and enhance overall well-being \cite{Brown2019}. Abdelkarim et al. found that enhancing infrastructure is essential by adding comfortable seating, shelters, lighting, and ensuring accessibility \cite{Abdelkarim2023}. Seat comfort, cleanliness inside the vehicle, and adequate lighting are key elements that can make the transport system more comfortable and enhance the usability, comfort, and overall appeal of public transportation spaces \cite{IMRE20172441}. Key improvements include infrastructure upgrades, technological enhancements, comfort and convenience features, and environmental sustainability initiatives \cite{Foth2013, DEONA2021129, Abdelkarim2023}. Similarly, Chin et al. approached public transportation space improvement from the user's perspective and conducted a survey on the users' key needs and preferences \cite{Chin2019}. The authors found that comfort, cleanliness, safety perception, accessibility, aesthetics, and technological integration are the key elements that positively impact passenger well-being.   

The environment within train cabins has crucial effects on shaping the experiences and well-being of passengers. Factors such as cleanliness, noise levels, lighting, and spatial design significantly influence passengers' comfort and satisfaction \cite{Shen2016}. For example, a clean environment has been shown to reduce stress and enhance mood, while a dirty or cluttered space can increase stress levels and negatively affect mental health \cite{Evans1998}. Similarly, perceived cleanliness is strongly correlated with passengers' willingness to use public transportation and their overall satisfaction with the service. According to \cite{Eboli2014}, perceived cleanliness, safety, and punctuality are crucial factors in rail passengers' satisfaction and service quality. Shen et al. have developed a comprehensive model for evaluating passenger satisfaction in urban rail transit systems using Structural Equation Modeling (SEM) based on Partial Least Squares (PLS) \cite{Shen2016}. Their research identifies key factors influencing passenger satisfaction, including perceived quality for punctuality, safety, cleanliness, comfort, perceived value, and passenger expectations.

Additionally, ambient conditions in the cabin such as the noise level can result in heightened stress, fatigue, and even long-term cardiovascular issues \cite{Stansfeld2003,Basner2014}. Another important factor is lighting. The lighting conditions within train cabins play a critical role in shaping passengers' mood and comfort. Natural lighting, in particular, has been found to enhance mood and well-being, making the incorporation of large windows in the design of train cabins a valuable feature \cite{Edwards2002,Veitch2001}. Studies have indicated that biophilic design elements, which integrate natural features into the built environment, can diminish stress levels while enhancing creativity and cognitive function \cite{Ulrich1991,Han2008,ARISTIZABAL2021101682}. In addition, incorporating design elements such as plants or artwork can cultivate a more pleasant and tranquil atmosphere, positively impacting passengers' mental well-being \cite{browning2014patterns,kellert2015practice}.



Lastly, from a productivity and space use point of view, the configuration of train cabins, including seating arrangements and available amenities, plays a crucial role in shaping passengers' experiences. Comfortable seating, ample legroom, and features like power outlets and Wi-Fi can elevate the journey, making it more enjoyable and conducive to productivity \cite{Hensher2003,Friman2009}. While several studies have examined the physical characteristics of public transportation cabins and possible enhancement alternatives, the influence of each of these improvement options on passengers' well-being measures and perceptions has not been compared to each other directly.

\subsection{Objective Perception Measurement Tool: Eye-tracking}

Eye-tracking technology measures where and how long a person looks at visual stimuli using infrared light and cameras to detect eye movement. It aids in understanding visual attention, perception, and cognitive processes\cite{holmqvist2011,duchowski2007}. Eye-tracking has various applications, including medical research\cite{Anderson2013}, education\cite{Hyn2010}, retail\cite{Grewal2009}, and transportation\cite{Tavakoli2022}. Recently, more affordable computer vision-based eye trackers using standard webcams have emerged, allowing for broader participant access despite lower accuracy\cite{Hutt2023}. Eye-tracking provides raw gaze vectors indicating fixation points and saccades, enabling the measurement of metrics like fixation count and duration, and saccade paths\cite{zhang2015,holmqvist2011}.

In built environments, eye-tracking studies have been utilized to understand how individuals interact visually with architectural spaces and urban landscapes. This data helps urban planners design better spaces by assessing which elements attract attention \cite{Han2022}. While this is still a new area of interest specific to train stations and public transportation spaces, a few studies have explored using this data modality for researching user experience. For instance, Schneider et al. used eye-tracking technology to determine users' experiences and their areas of interest in a crowded train station \cite{Schneider2023}. The authors found out that the signalization prototype intended to regulate pedestrian flow was visible in low crowd density conditions but less effective in high crowd density, where herd behavior influenced path choice more significantly. Additionally, Höller et al. used eye-tracking to investigate how people perceive information screens in public transportation stations and how much time it takes for them to receive the information they need \cite{Holler2021}. The authors ran an exploratory field study with 106 participants and the results indicated a high awareness of the info-screens among participants but found no significant correlations between fixation time and content type or between fixation time and recall/recognition of content.

\section{Methodology}

In this study, participants were asked to answer a series of standard questionnaires about their well-being (i.e., emotional valence and arousal, stress level, and creativity) and perceptual metrics (i.e., reasonable price, comfort level, safety, cleanliness, passage of time, likelihood to recommend to others, utilization of workspace in the cabin, and punctuality) after viewing pictures of various cabin designs. Their eye gaze pattern was monitored while they looked at the pictures using an online eye tracking service.  

\subsection{Image Selection}
\begin{figure*}
	\centering
	\includegraphics[width=1\textwidth]{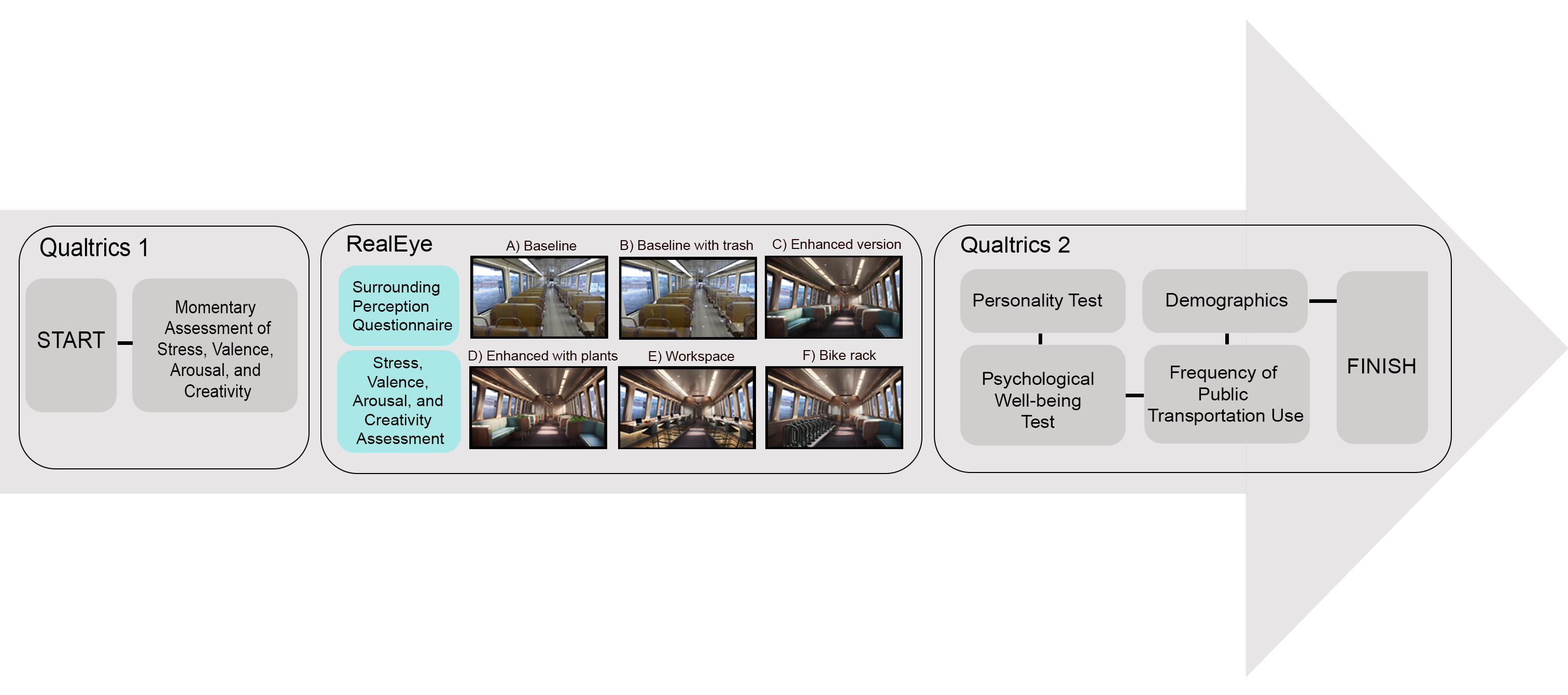}
	\caption{The study design framework is depicted, with blue representing the RealEye platform and gray representing Qualtrics. Images were displayed randomly within RealEye.}
	\label{procedure}
\end{figure*}

The images in this study were created using extensive graphic editing and rendering with the accommodations provided by advanced Artificial Intelligence (AI) technology, such as GPT-4 (DALL-E). These methods were used to produce a range of designs for the cabins of the same train. The images made it possible to consistently and accurately visualize different cabin layouts, enabling a controlled comparison of how different design elements affect passenger perception and well-being.
The images depicted a range of cabin configurations including (1) \textit{the current version} of the cabin space as a baseline taken from a regional train in the United States, (2) \textit{the current version with added trash and visible wear and tear} as an example of lack of maintenance, (3) \textit{an aesthetically enhanced version} of the current cabin space which involved changing the material, seat arrangement, and presentation, (4) a \textit{bike rack enabled} version featuring a bike rack as an example for a design that accommodates multimodal transportation needs, (5) another with an \textit{added workspace} as an example that accommodates personal and work-related needs, and lastly, (6) \textit{an enhanced version with added plants} an example of having biophilic designs (Fig. \ref{procedure}).

\subsection{Study Design and Flow}

The following survey was designed as a within-subject study. The study involved using two Qualtrics Surveys and an online eye-tracking procedure performed by the RealEye platform \cite{RealEye}.

In the first Qualtrics survey, participants responded to the consent form and were asked about their current state of well-being (stress, valence, arousal, and creativity) to establish a baseline. Participants were then moved to the RealEye environment for eye-tracking. The first task in the RealEye environment was calibration, which is a standard procedure administered by the RealEye website \cite{RealEye}. Once calibration was complete, participants were shown a series of six different images presented in a random order to eliminate bias. Before each image, participants were instructed to imagine it as a cabin on their daily commute train with the prompt: \textit{"Imagine this is a cabin in the train you use for your daily commute"}.  While viewing the images in random order, participants’ eye gaze patterns were recorded through the RealEye platform. 

After each image, participants answered a set of questions related to the study's dependent variables, including well-being metrics (stress, valence, arousal, and creativity) and surrounding perception metrics (desired price, comfort level, safety, cleanliness, passage of time, likelihood to recommend to others, utilization of workspace in the cabin, and punctuality). Participants then viewed a black screen for 3 seconds before moving on to the next image. Once all six images had been viewed, participants were directed to the second Qualtrics survey where they responded to questions about their baseline demographics, and the frequency of their use of public transportation.

\subsubsection{Well-being Metrics: Momentary Stress, Valence, Arousal, and Creativity}

To assess well-being we used established questionnaires from the literature to measure momentary stress, emotion (i.e., arousal, and valence), and creativity levels. Stress was measured using a 0-10 Likert scale with the question "\textit{What describes your level of stress right now?}" based on \cite{Karvounides2016-ja}. Valence was assessed on a 7-point Likert scale by asking "\textit{How positive, neutral, or negative do you feel right now?}"; arousal was measured on a 7-point Likert scale with the question "\textit{How aroused do you feel right now? Aroused means activated, charged, alert, or energized}", both from \cite{Mauss2009-sr}. Creativity was gauged by asking "\textit{How creative do you feel in this cabin?}" on a 5-point scale from "not creative at all" to "very creative," adapted from \cite{Kaufman2012}.
 
\subsubsection{Perceptual Metrics}

To evaluate how participants perceive each cabin’s built environment, we concentrated on eight specific aspects: the passage of time, safety, cleanliness, comfort, punctuality of trains, reasonable cost associated with it, likelihood to recommend to others, and utilization of workspace in the cabin. Some of the questions were adopted from prior literature, while others were created specifically for this study. The questions are as follows:
\begin{itemize}
\item Cleanliness:
\textit{Regarding the cabin you just observed, please indicate your agreement/disagreement with the following statement: “This cabin is clean.”} on a 5-point scale from fully disagree to fully agree.
\item Comfort:
\textit{Regarding the cabin you just observed, please indicate your agreement/disagreement with the following statement: “Traveling in this cabin is comfortable.”} on a 5-point scale from fully disagree to fully agree.
\item  Safety:
\textit{Please rate how safe or unsafe you feel in this cabin.} on a 5-point scale from completely unsafe to completely safe, adopted from \cite{PTC2023}
\item  Passage of time:
\textit{How would you rate the passage of time in this cabin?} on a 5-point scale from very slow to very fast, adopted from \cite{altaf2023time}.
\item  Price:
\textit{If an average train ticket costs you \$5, how much are you willing to pay to use this train?} on a 9-point scale from \$1 to \$9, adopted from \cite{Wertenbroch2002}.
\item  Punctuality:
\textit{Regarding the cabin you just observed, please indicate your agreement/disagreement with the following statement: “The train that includes this cabin mostly runs on schedule.”} on a 5-point scale from fully disagree to fully agree.
\item  Likelihood to recommend to others: \textit{Could you please indicate the extent to which you agree with the following statement? “I would recommend this train to other passengers.”} on a 5-point scale from fully disagree to fully agree, adopted from \cite{Reichheld2003}.
\item  Workspace utility:
\textit{Please imagine yourself commuting on this train to your workplace. How much would you want to utilize this cabin as a workspace during your journey to work?} on a 5-point scale from very unlikely to very likely.
\end{itemize}
\subsubsection{Personality, and Demographic Questions}

We gathered information about participants' personality traits using a shortened version of the Big Five Inventory \cite{Gosling2003} as well as demographics such as ethnicity, age, and gender.

\begin{table}[h!]
\centering
\caption{Demographics of participants (N=304)}\label{demographics}
\begin{tabular}{llcc}
\hline
Variable & Attribute & Number & Percent \\
\hline
\multicolumn{4}{c}{\bfseries} \\
Gender & Non-male & 149 & 49.01 \\
 & Male & 155 & 50.98 \\
\hline
\multicolumn{4}{c}{\bfseries} \\
Ethnicity & White & 213 & 70.06 \\
 & Black or African American & 41 & 13.48 \\
 & American Indian or Alaska Native & 1 & 0.03 \\
 & Asian & 34 & 11.18 \\
 & Other & 12 & 3.94 \\
 & Prefer not to say & 3 & 1.0 \\
\hline
\multicolumn{4}{c}{\bfseries} \\
Education & Less than a Bachelor’s degree & 80 & 26.31 \\
 & Bachelor’s degree or higher & 223 & 73.35 \\
 & Prefer not to say & 1 & 0.3 \\
\hline
\multicolumn{4}{c}{\bfseries} \\
Age & Mean (Std. Dev) & 39.60 (13.45) & \\
& Range & 19--78 \\
\hline
\end{tabular}
\end{table}

\subsection{Participant Recruitment}

The study was approved by the Villanova University Institutional Review Board (IRB) and is documented in the IRB \#FY2024-92. Participants from the Northeastern region of the United States were recruited through the Prolific Platform \cite{prolific}. Each participant received \$5.20 as compensation for their 15-minute participation in the study. A total of 304 participants completed the study. Table \ref{demographics} provides details of participants' gender, ethnicity, education, and age distribution.

\subsection{Eye-tracking Software}

The RealEye platform \cite{RealEye} was used to track users' gaze patterns with their own laptop cameras. While the adoption of RealEye is a recent advancement in this arena of research, other studies have utilized this platform for similar studies such as \cite{10.1145/3517031.3529615,FEDERICO2019103582}. RealEye identifies the specific locations where participants focus their gaze, referred to as fixation locations, the duration of each instance of focus, referred to as fixation duration, and the time and gaze path between fixation points referred to as saccadic movements, for every participant. RealEye employs default settings to assess fixation, which includes a 200 ms noise reduction window, an 80 ms minimum fixation time, a 400 ms maximum fixation length, and a 150 ms maximum saccade duration. After finishing the eye tracking, a CSV file containing this data is exported to extract further features. For brevity, in this article, we will only focus on a preliminary analysis of the eye-tracking data.  

 





\subsection{Statistical Modeling}
In our analysis of the statistical variations in the dependent variables (e.g., momentary stress level) across the six conditions, we utilized linear mixed-effect models (LMM) \cite{baayen2008mixed}. The choice of LMMs was deliberate, as they are capable of accommodating the interparticipant dependence inherent in our within-subject study design. This enables LMMs to effectively handle the repeated measures from the same participants, acknowledging that measurements from the same individual are not independent of each other. This versatility enables the models to adjust for individual differences in baseline measurements and other random fluctuations that may influence the dependent variables. It should be noted that for the well-being metrics, a baseline assessment was included as a pre-experiment to the conditions.

We utilized the lme4 package in the R programming language for our modeling \cite{bates2015fitting}. Widely embraced for training and fitting linear and nonlinear mixed-effects models, the lme4 package provided us with a robust and adaptable framework for analyzing the specific data structures commonly encountered in psychological research, which is often complicated through random factors introduced by having different baselines for various participants.

\section{Results}

In the following sections, we will present the results of each LMM model developed for each of the dependent variables. Please note that in the following sections, CI stands for confidence interval, df stands for degrees of freedom, and Pr stands for the probability of being greater than the t-value test statistic from the mixed effect model.

\subsection{Well-being Metrics}
\subsubsection{Perceived Stress}

\begin{figure}[!h]
	\centering
	\includegraphics[width=0.47\textwidth]{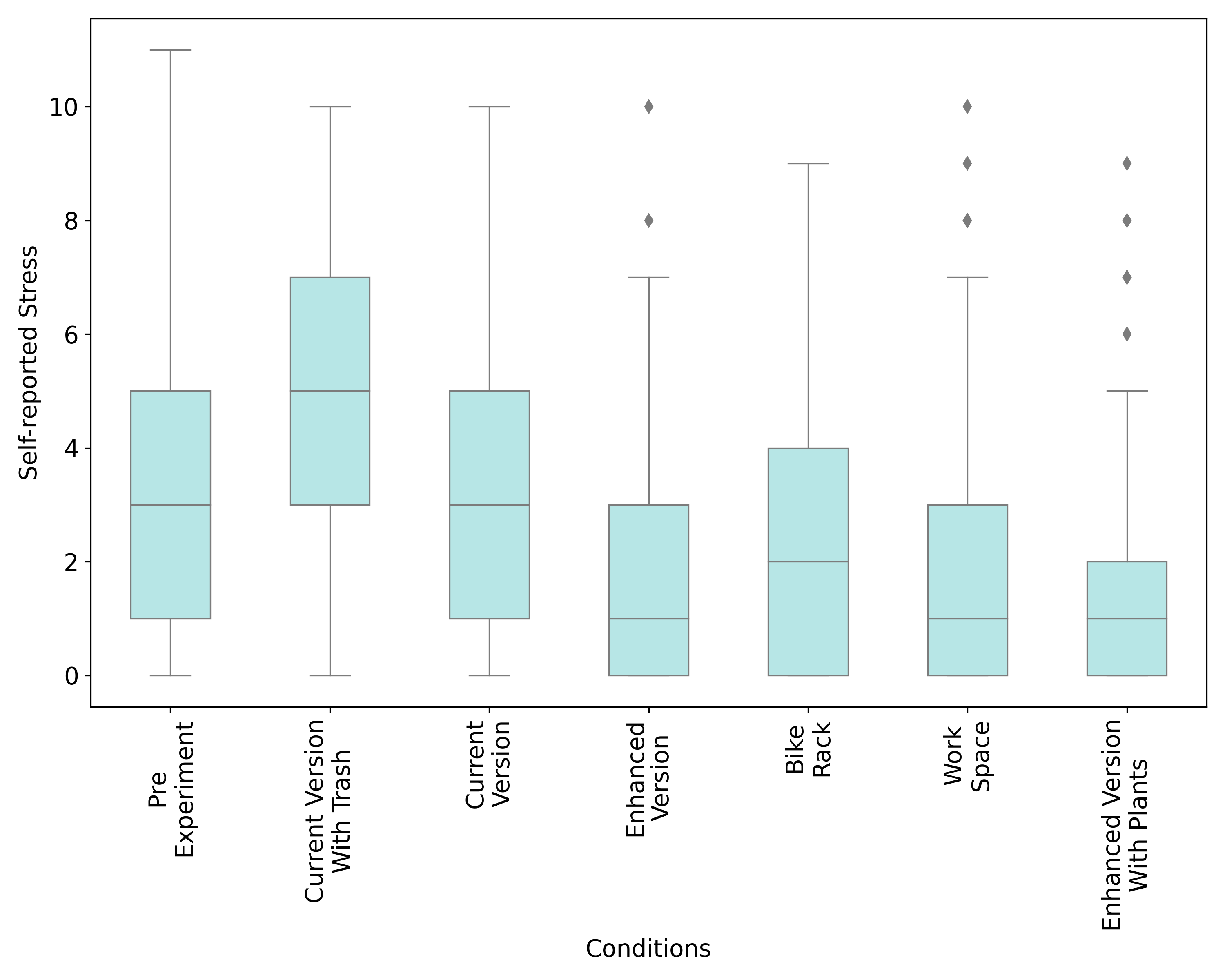}
	\caption{Comparison of perceived stress across different conditions}
	\label{stress}
\end{figure}

The stress levels of participants varied significantly across all seven conditions of pre-experiment, enhanced version, enhanced version with plants, cabin containing bike racks, cabin containing workspace, and the current version with added trash when compared to the current condition of the public transportation built environment, referred to as \textit{current version} as shown in Fig. \ref{stress}. Additionally, the model indicates that the cabin space with plants has the highest absolute value of $\beta$ estimate for the model ($\beta$ = -1.741). The $\beta$ estimates for workspace, bike rack, and enhanced version are all significant and negative, suggesting a decrease in stress levels under those conditions. Conversely, the current version with added trash has the highest positive $\beta$ ($\beta$ = 1.834), indicating that the presence of trash in transportation spaces increases stress as shown in Table \ref{stress}.

\begin{table}[!h]
\caption {The effect of different transportation space conditions on participants’ stress
level, evaluated with respect to the current version.}\label{stress_table} 
\resizebox{0.47\textwidth}{!}{%
\begin{tabular}{lrrrrrrr}
Variable                     & \multicolumn{1}{l}{Estimate} & \multicolumn{1}{l}{std.error} & \multicolumn{1}{l}{t-value} & \multicolumn{1}{l}{df} & \multicolumn{1}{l}{Pr(\textgreater{}|t|)} & \multicolumn{1}{l}{2.5 \% CI} & \multicolumn{1}{l}{97.5 \% CI} \\ \hline
Current Version (Intercept)  & 2.596                        & 0.406                         & 6.398                       & 326.04                 & 5.45E-10                                 & 1.798                         & 3.394                          \\
Current Version With Trash   & 1.834                        & 0.130                         & 14.085                      & 1800                   & 8.10E-43                                 & 1.579                         & 2.089                          \\
Bike Rack                    & -0.728                       & 0.130                         & -5.588                      & 1800                   & 2.65E-08                                 & -0.983                        & -0.472                         \\
Pre-experiment               & 0.226                        & 0.130                         & 1.735                       & 1800                   & 8.29E-02                                 & -0.029                        & 0.481                          \\
Work Space                   & -1.093                       & 0.130                         & -8.395                      & 1800                   & 9.32E-17                                 & -1.348                        & -0.838                         \\
Enhanced Version             & -1.502                       & 0.130                         & -11.533                     & 1800                   & 9.68E-30                                 & -1.757                        & -1.246                         \\
Enhanced Version With Plants & -1.741                       & 0.130                         & -13.370                     & 1800                   & 6.09E-39                                 & -1.996                        & -1.485                         \\
Gender                       & 0.238                        & 0.198                         & 1.201                       & 298                    & 2.31E-01                                 & -0.152                        & 0.628                          \\
Ethnicity                    & 0.104                        & 0.216                         & 0.481                       & 298                    & 6.31E-01                                 & -0.322                        & 0.530                          \\
\end{tabular}%
}
\end{table}

\subsubsection{Valence}
\begin{figure}[!h]
	\centering
	\includegraphics[width=0.47\textwidth]{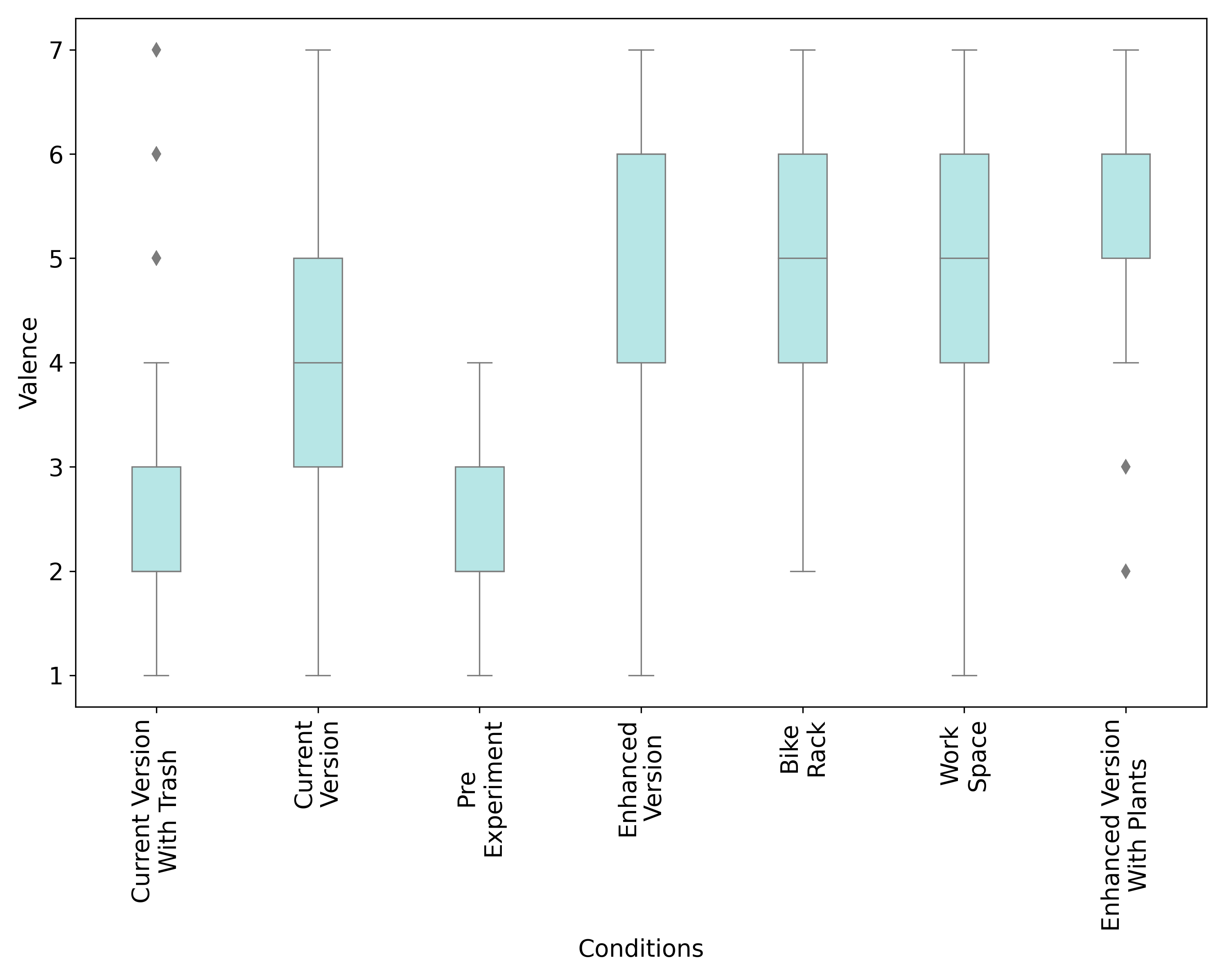}
	\caption{Comparison of emotional valence across different conditions}
	\label{valence}
\end{figure}

As shown in Fig.\ref{valence}, participants experienced different levels of emotional valence after viewing each image of the enhanced version, the enhanced version with plants, bike racks, workspace in the cabin, and the current version with added trash when compared to the existing state of the public transportation built environment referred to as current version, as depicted in Table \ref{valence table}. The presence of added trash in the environment was associated with the most pronounced decrease in valence towards negative emotions ($\beta$=-1.229). Conversely, the enhanced version with plants had the highest estimated value ($\beta$=1.831), indicating an increase in positive valence.  

\begin{table}[!h]
\caption{The effect of different transportation space conditions on participants’ valence
level, evaluated with respect to the current version.}\label{valence table} 
\resizebox{0.47\textwidth}{!}{%
\begin{tabular}{llllllll}
Variable                     & Estimate & std.error  & t-value & df     & Pr(\textgreater{}|t|) & 2.5 \% CI & 97.5 \% CI \\ \hline
Current Version (Intercept)  & −0.203   & 0.16718772 & -1.2124 & 389.37 & 2.26E-01              & −0.531    & 0.126      \\
Current Version With Trash   & −1.229   & 0.0908978  & -13.523 & 1800   & 9.29E-40              & −1.408    & −1.051     \\
Bike Rack                    & 0.99     & 0.0908978  & 10.892  & 1800   & 8.48E-27              & 0.812     & 1.168      \\
Pre\_Experiment              & −1.326   & 0.0908978  & -14.583 & 1800   & 1.29E-45              & −1.504    & −1.147     \\
Work Space                   & 1.458    & 0.0908978  & 16.045  & 1800   & 2.97E-54              & 1.28      & 1.637      \\
Enhanced Version             & 1.498    & 0.0908978  & 16.484  & 1800   & 5.78E-57              & 1.32      & 1.677      \\
Enhanced Version With Plants & 1.831    & 0.0908978  & 20.139  & 1800   & 1.65E-81              & 1.652     & 2.009      \\
Gender                       & −0.122   & 0.07804055 & -1.5674 & 298    & 1.18E-01              & −0.276    & 0.031      \\
Ethnicity                    & 0.166    & 0.08521861 & 1.9489  & 298    & 5.23E-02              & −0.002    & 0.334     
\end{tabular}%
}
\end{table}

\subsubsection{Arousal} 
\begin{figure}[!h]
	\centering
	\includegraphics[width=0.47\textwidth]{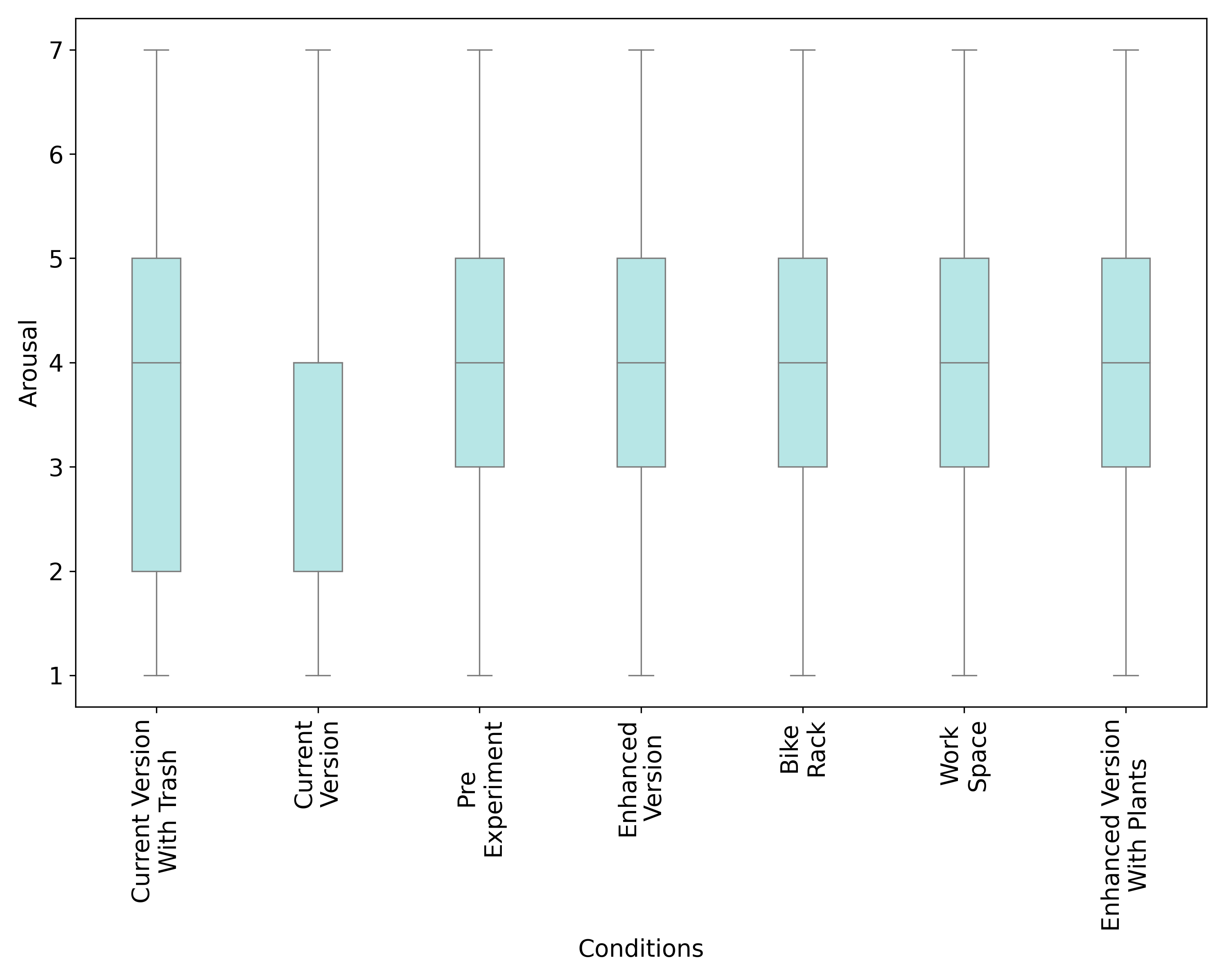}
	\caption{Comparison of arousal across different conditions}
	\label{arousal}
\end{figure}

Participants' arousal did not change significantly across different experimental conditions, as depicted in Fig. \ref{arousal}. However, the current version differed from all other conditions. The disparity in arousal is evident in the workspace ($\beta$=0.963) and the enhanced version with plants ($\beta$=0.721), as shown in Table \ref{arousal table}. This indicates an increase in arousal levels compared to the current version of the transportation space.

\begin{table}[!h]
\caption{The effect of different transportation space conditions on participants’ arousal
level, evaluated with respect to the current version.}\label{arousal table}  
\resizebox{0.47\textwidth}{!}{%
\begin{tabular}{llllllll}
Variable                     & Estimate & std.error & t-value & df     & Pr(>|t|) & 2.5 \% CI & 97.5 \% CI \\ \hline
Current Version (Intercept)  & 3.272    & 0.3036    & 10.779  & 326.01 & 2.25E-23 & 2.675    & 3.87      \\
Current Version With Trash   & 0.176    & 0.0974    & 1.8082  & 1800   & 7.07E-02 & −0.015   & 0.367     \\
Bike Rack                    & 0.359    & 0.0974    & 3.6845  & 1800   & 2.36E-04 & 0.168    & 0.55      \\
Pre\_Experiment               & 0.512    & 0.0974    & 5.2539  & 1800   & 1.67E-07 & 0.321    & 0.703     \\
Work Space                   & 0.963    & 0.0974    & 9.8937  & 1800   & 1.64E-22 & 0.772    & 1.154     \\
Enhanced Version             & 0.392    & 0.0974    & 4.0257  & 1800   & 5.92E-05 & 0.201    & 0.583     \\
Enhanced Version With Plants & 0.721    & 0.0974    & 7.4032  & 1800   & 2.03E-13 & 0.53     & 0.912     \\
Gender                       & −0.083   & 0.1483    & -0.5623 & 298    & 5.74E-01 & −0.375   & 0.208     \\
Ethnicity                    & 0.137    & 0.1619    & 0.844   & 298    & 3.99E-01 & −0.182   & 0.455    
\end{tabular}%
}
\end{table}

\subsubsection{Creativity}
\begin{figure}[!h]
	\centering
	\includegraphics[width=0.47\textwidth]{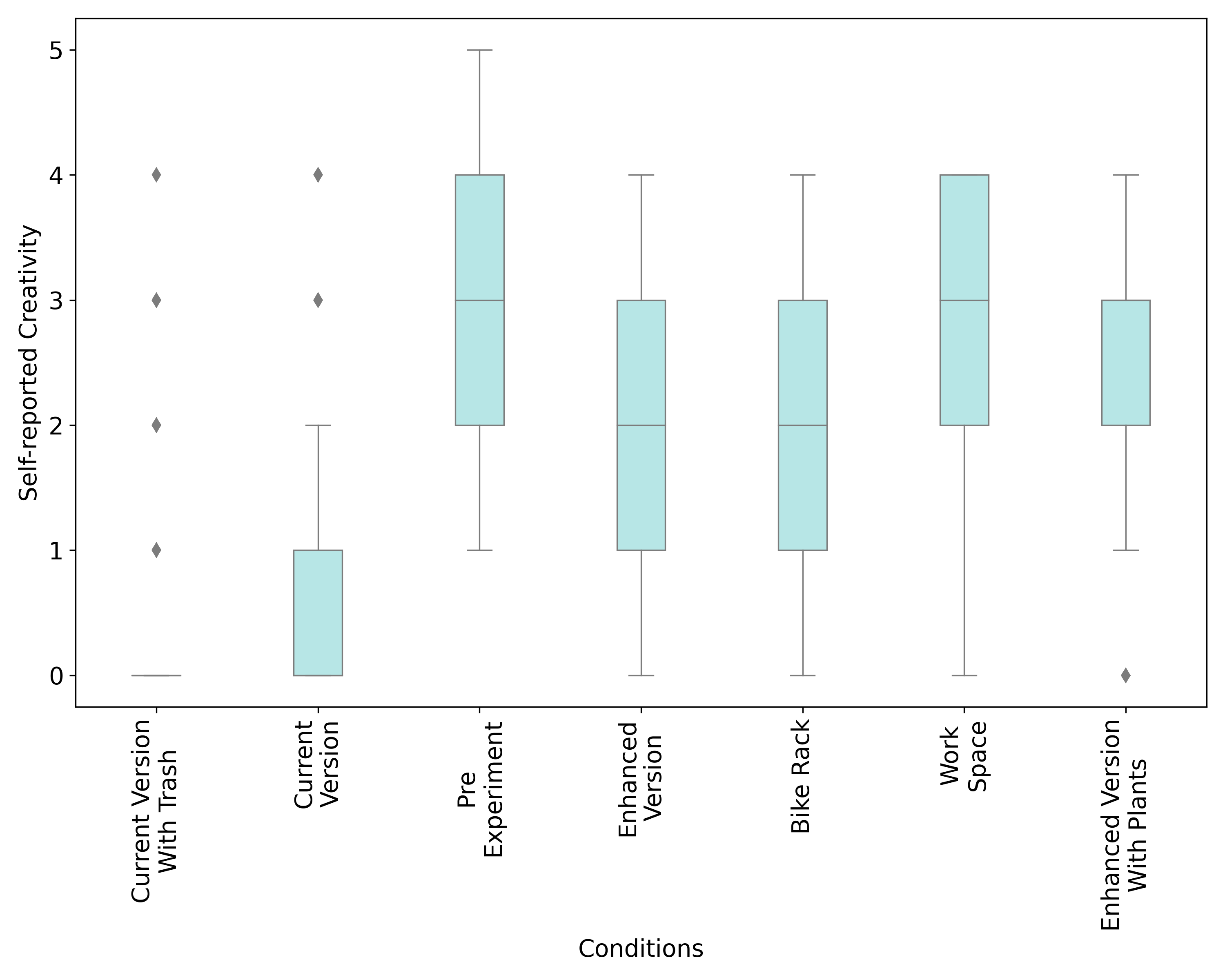}
	\caption{Comparison of creativity across different conditions}
	\label{creativity}
\end{figure}

Similar to valence and stress, participant creativity has been significantly different in all six conditions of pre-experiment, enhanced version, enhanced version with plants, Bike racks, the workspace, and presence of trash compared to the current version of public transportation interior space as depicted in Fig. \ref{creativity}. Estimated $\beta$ for the different conditions implies that the enhanced version with plants ($\beta$=1.814) and the workspace ($\beta$=1.814) had a positive impact on creativity compared to the current version. The conditions with bike racks in the cabin and the enhanced version are also significantly improving creativity (Table \ref{Creativity table}. Additionally, participants' gender has a significant $\beta$ value, indicating an impact on the creativity level with non-males reported scoring approximately 0.243 points lower in perceived creativity in the imagined cabins compared to males.   

\begin{table}[!h]
\caption {The effect of different transportation space conditions on participants’ creativity
level, evaluated with respect to the current version.} \label{Creativity table} 
\resizebox{0.47\textwidth}{!}{%
\begin{tabular}{llllllll}
Variable                               & Estimate & std.error & t-value & df     & Pr(\textgreater{}|t|) & 2.5 \% CI & 97.5 \% CI \\ \hline
Current Version (Intercept) & 0.956    & 0.1707    & 5.5986  & 354.83 & 4.33E-08              & 0.62      & 1.292      \\
Current Version With Trash             & −0.425   & 0.0757    & -5.6186 & 1800   & 2.23E-08              & −0.574    & −0.277     \\
Bike Rack                              & 0.924    & 0.0757    & 12.203  & 1800   & 5.78E-33              & 0.775     & 1.072      \\
Pre-experiment                         & 2.003    & 0.0757    & 26.469  & 1800   & 1.13E-130             & 1.855     & 2.152      \\
Work Space                             & 1.814    & 0.0757    & 23.967  & 1800   & 2.12E-110             & 1.666     & 1.962      \\
Enhanced Version                       & 1.385    & 0.0757    & 18.304  & 1800   & 8.98E-69              & 1.237     & 1.534      \\
Enhanced Version With Plants           & 1.814    & 0.0757    & 23.967  & 1800   & 2.12E-110             & 1.666     & 1.962      \\
Gender                                 & −0.243   & 0.0816    & -2.9733 & 298    & 3.19E-03              & −0.403    & −0.082     \\
Ethnicity                              & 0.126    & 0.0891    & 1.4128  & 298    & 1.59E-01              & −0.049    & 0.301     
\end{tabular}%
}
\end{table}

\subsection{Transportation space perception metrics}
\subsubsection{Cleanliness}
\begin{figure}[!h]
	\centering
	\includegraphics[width=0.47\textwidth]{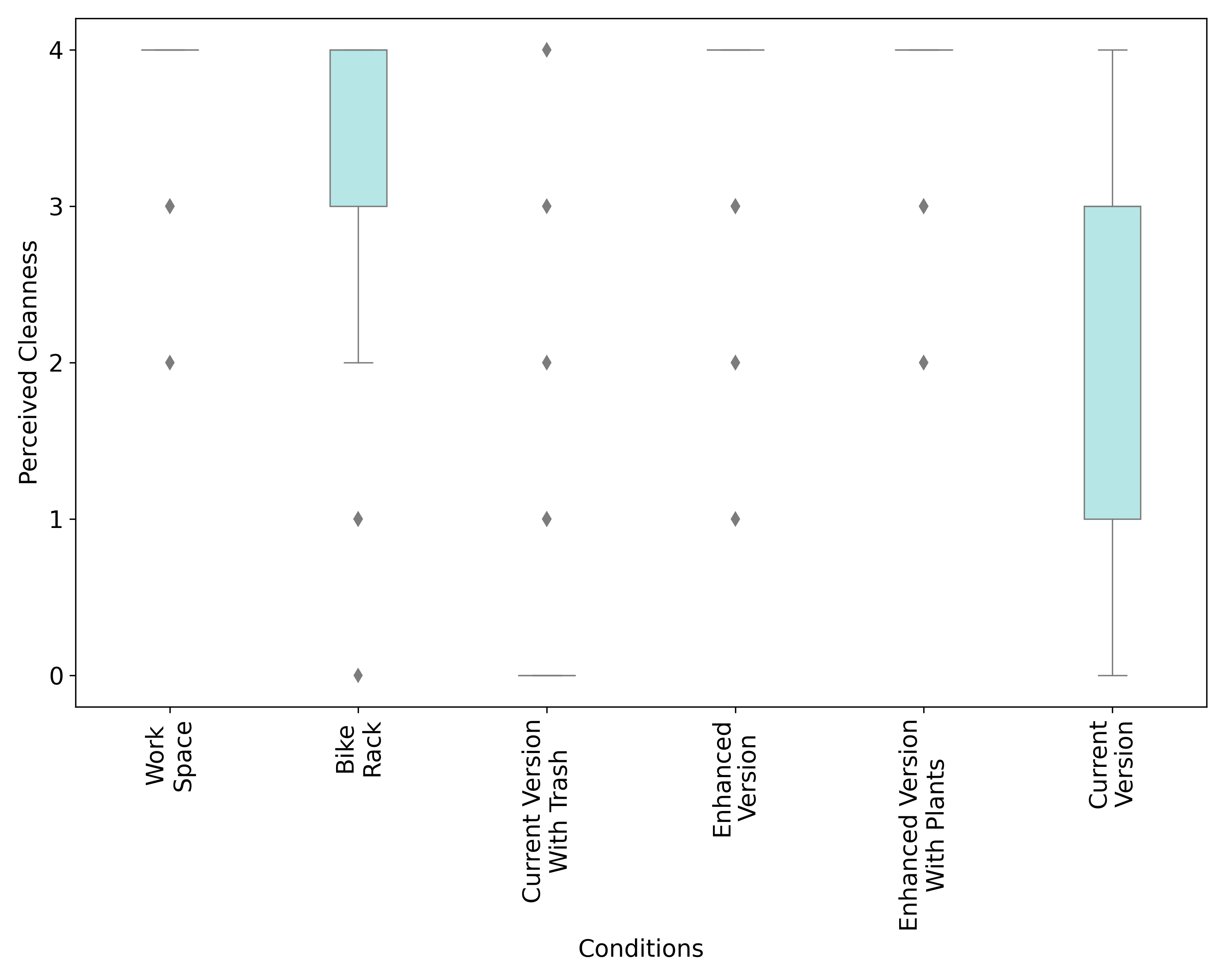}
	\caption{Comparison of perceived cleanliness across different conditions}
	\label{cleanness}
\end{figure}

The perceived cleanliness by participants was significantly different between the five conditions of enhanced version, enhanced version with plants, Bike racks, added workspace, and current version with added trash compared to a current version of transportation space as depicted in Fig. \ref{cleanness}. The current version with trash had the highest absolute value estimated for $\beta$ ($\beta$=-2.172), indicating that the presence of trash decreases the perception of cleanliness compared to the current version of public transportation space. On the other hand, Workspace has the highest positive $\beta$ ($\beta$=1.454) indicating an increase in perceived cleanliness in comparison to the current version (Table \ref{cleanness table}). 

\begin{table}[!h]
\caption {The effect of different designs of transportation spaces on participants’ perception of the built environment's cleanliness
level, evaluated with respect to the current version.} \label{cleanness table}
\resizebox{0.47\textwidth}{!}{%
\begin{tabular}{llllllll}
Variable                     & Estimate & std.error & t-value & df     & Pr(\textgreater{}|t|) & 2.5 \% CI & 97.5 \% CI \\ \hline
Current Version (Intercept)                  & 0.298    & 0.1006    & 2.9576  & 395.26 & 3.29E-03              & 0.1       & 0.496      \\
Current Version With Trash   & −2.172   & 0.0567    & -38.308 & 1505   & 1.09E-224             & −2.283    & −2.061     \\
Bike Rack                    & 0.815    & 0.0567    & 14.366  & 1505   & 5.96E-44              & 0.703     & 0.926      \\
Work Space                   & 1.454    & 0.0567    & 25.636  & 1505   & 1.43E-120             & 1.342     & 1.565      \\
Enhanced Version             & 1.328    & 0.0567    & 23.417  & 1505   & 1.16E-103             & 1.217     & 1.439      \\
Enhanced Version With Plants & 1.394    & 0.0567    & 24.585  & 1505   & 1.76E-112             & 1.283     & 1.505      \\
Gender                       & 0.081    & 0.0467    & 1.7298  & 299    & 8.47E-02              & −0.011    & 0.173      \\
Ethnicity                    & −0.006   & 0.0509    & -0.1134 & 299    & 9.10E-01              & −0.106    & 0.094     
\end{tabular}%
}
\end{table}

\subsubsection{Perceived Comfort}
\begin{figure}[!h]
	\centering
	\includegraphics[width=0.47\textwidth]{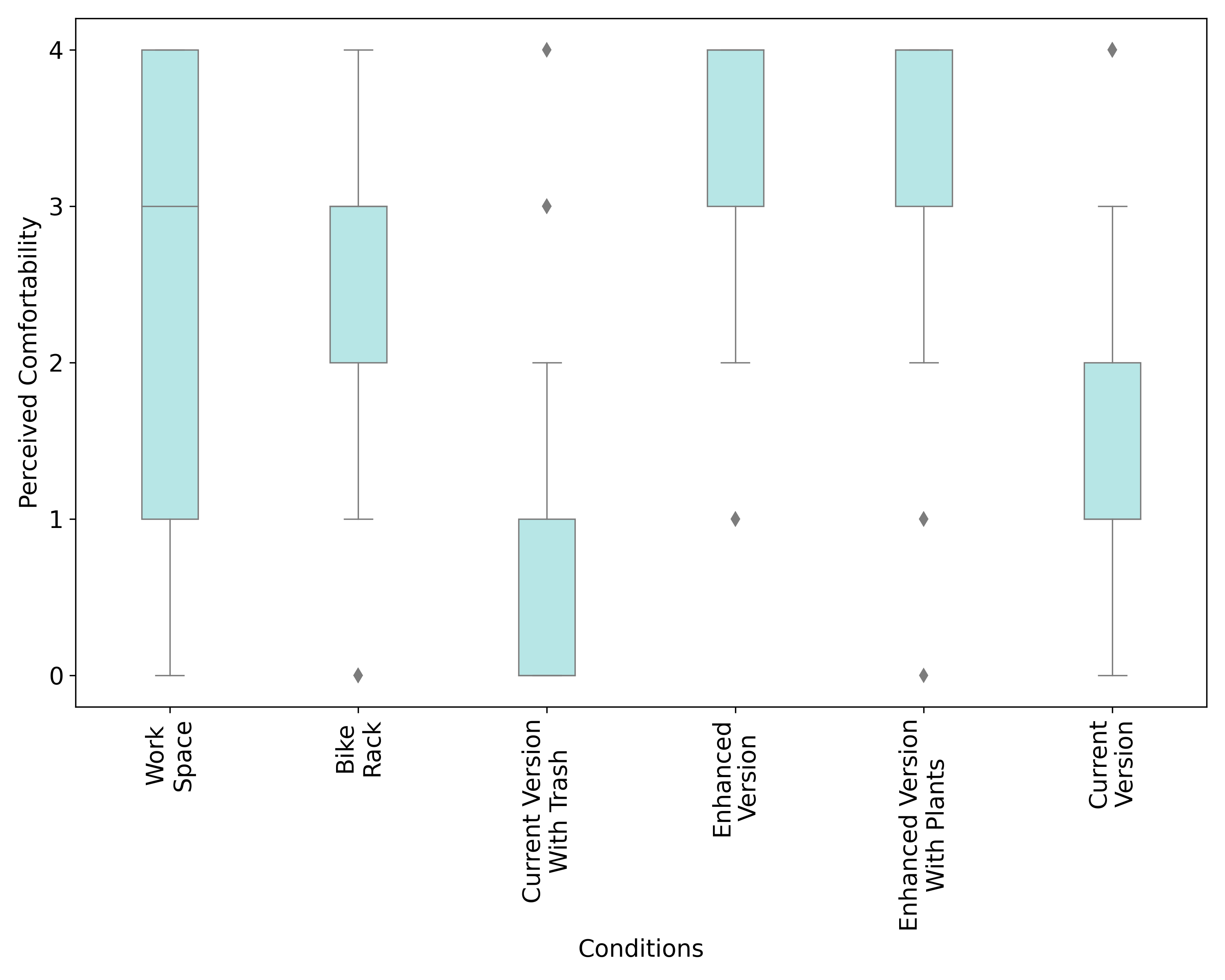}
	\caption{Comparison of perceived comfort level across different conditions}
	\label{comfort}
\end{figure}

As depicted in Fig.\ref{comfort} participants reported significantly different perceived comfort levels across various conditions, including the enhanced version, enhanced version with plants, bike racks, workspace, and current version with added trash, compared to the current version of transportation space. As shown in Table \ref{comfort table} the enhanced version with plants($\beta$=1.894), the enhanced version ($\beta$= 1.775), Workspace ($\beta$=0.954), and bike rack ($\beta$=0.904) significantly increases perceptions of comfort in comparison to the current version of the public transportation. Also, adding trash decreases the perception of comfort in the built environment of the trains ($\beta$= -1.043) compared to the current version.

\begin{table}[!h]
\caption {The effect of different transportation space conditions on participants’ perception of comfort in the built environment, evaluated with respect to the current version.}\label{comfort table} 
\resizebox{0.47\textwidth}{!}{%
\begin{tabular}{lrrrrrrr}
Variable                     & Estimate & std.error & t-value & df     & Pr(\textgreater{}|t|) & 2.5 \% CI & 97.5 \% CI \\ \hline
Current Version (Intercept)  & -0.310                       & 0.137                         & -2.264                      & 393.71                 & 2.41E-02                                 & -0.579                        & -0.041                         \\
Current Version With Trash   & -1.043                       & 0.077                         & -13.625                     & 1505                   & 5.97E-40                                 & -1.193                        & -0.893                         \\
Bike Rack                    & 0.904                        & 0.077                         & 11.809                      & 1505                   & 7.71E-31                                 & 0.754                         & 1.054                          \\
Work Space                   & 0.954                        & 0.077                         & 12.458                      & 1505                   & 5.65E-34                                 & 0.803                         & 1.104                          \\
Enhanced Version             & 1.775                        & 0.077                         & 23.185                      & 1505                   & 6.24E-102                                & 1.625                         & 1.925                          \\
Enhanced Version With Plants & 1.894                        & 0.077                         & 24.742                      & 1505                   & 1.11E-113                                & 1.744                         & 2.044                          \\
Gender                & -0.108                       & 0.064                         & -1.696                      & 299                    & 9.08E-02                                 & -0.233                        & 0.017                          \\
Ethnicity             & 0.087                        & 0.069                         & 1.251                       & 299                    & 2.12E-01                                 & -0.050                        & 0.223                          \\
\end{tabular}%
}
\end{table}

\subsubsection{Safety}
\begin{figure}[!h]
	\centering
	\includegraphics[width=0.47\textwidth]{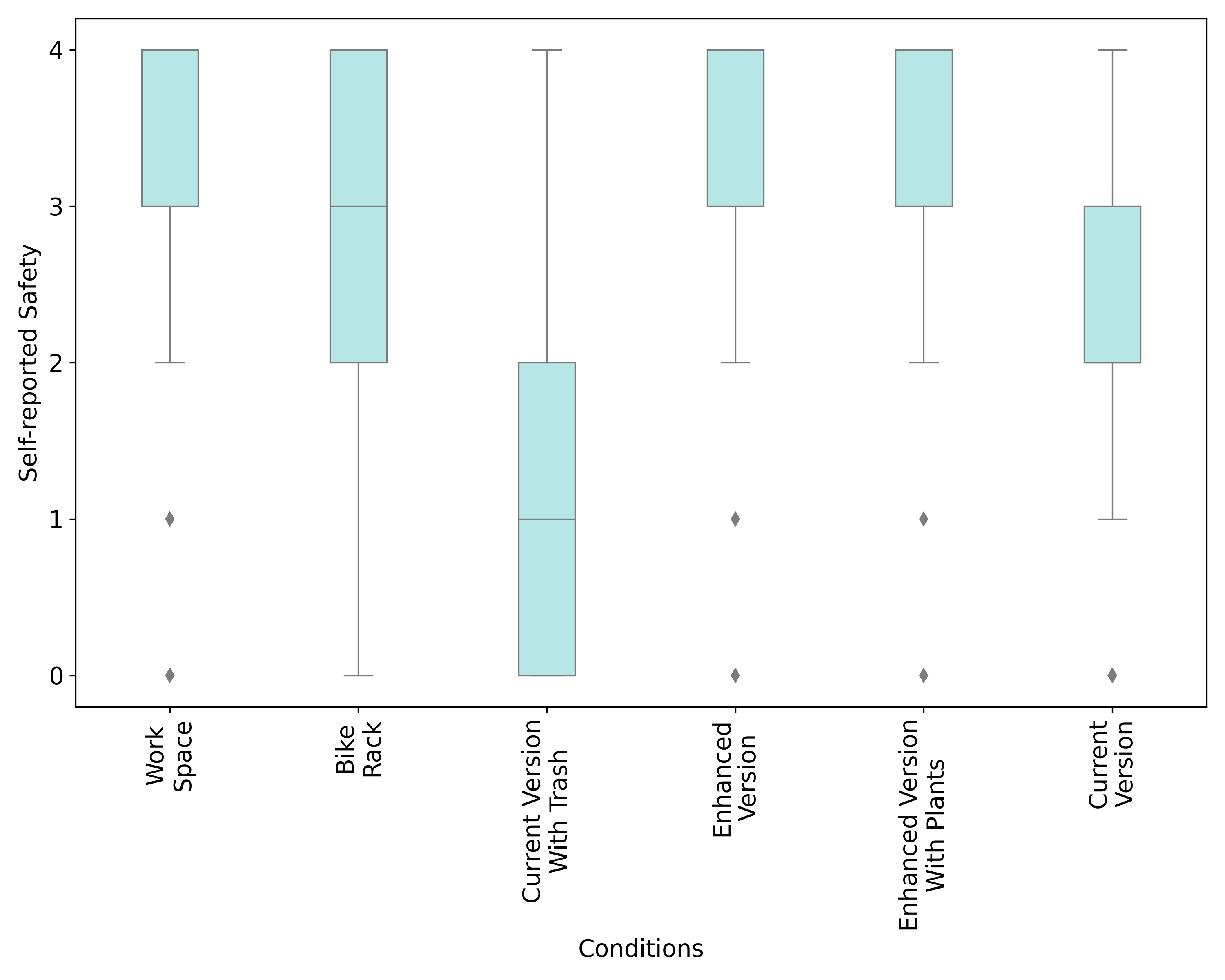}
	\caption{Comparison of perceived safety level across different conditions}
	\label{safety}
\end{figure}

Participants’ safety perceptions significantly varied across five conditions: enhanced version, enhanced version with plants, bike racks, workspace, and current version with added trash, compared to the current version, as depicted in Fig. \ref{safety}. As shown in Table \ref{Safety table}, among all the conditions, the $\beta$ estimate for participants' perception of safety was highest in the enhanced version with plants condition ($\beta$=1.212), indicating that this space increases safety perception compared to a standard cabin environment. Conversely, the current version with trash has the most negative estimate for $\beta$, ($\beta$= -1.185) indicating a decrease in participants' perception of safety.

\begin{table}[!h]
\caption {The effect of different public transportation interior space conditions on participants’ safety
perception, evaluated with respect to the current version.}\label{Safety table} 
\resizebox{0.47\textwidth}{!}{%
\begin{tabular}{llllllll}
Variable                     & Estimate & std.error & t-value & df     & Pr(\textgreater{}|t|) & 2.5 \% CI & 97.5 \% CI \\ \hline
Current Version (Intercept)                  & 0.671    & 0.1345    & 4.9894  & 362.35 & 9.42E-07              & 0.407     & 0.936      \\
Current Version With Trash   & −1.185   & 0.0634    & -18.699 & 1505   & 2.52E-70              & −1.310    & −1.061     \\
Bike Rack                    & 0.662    & 0.0634    & 10.447  & 1505   & 1.04E-24              & 0.538     & 0.787      \\
Work Space                   & 0.993    & 0.0634    & 15.67   & 1505   & 2.22E-51              & 0.869     & 1.118      \\
Enhanced Version             & 1.089    & 0.0634    & 17.185  & 1505   & 1.42E-60              & 0.965     & 1.214      \\
Enhanced Version With Plants & 1.212    & 0.0634    & 19.117  & 1505   & 4.17E-73              & 1.088     & 1.336      \\
Gender                       & −0.205   & 0.0639    & -3.2061 & 299    & 1.49E-03              & −0.330    & −0.079     \\
Ethnicity                    & −0.042   & 0.0696    & -0.6051 & 299    & 5.46E-01              & −0.179    & 0.095     
\end{tabular}%
}
\end{table}

\subsubsection{Passage of time}

\begin{figure}[!h]
	\centering
	\includegraphics[width=0.47\textwidth]{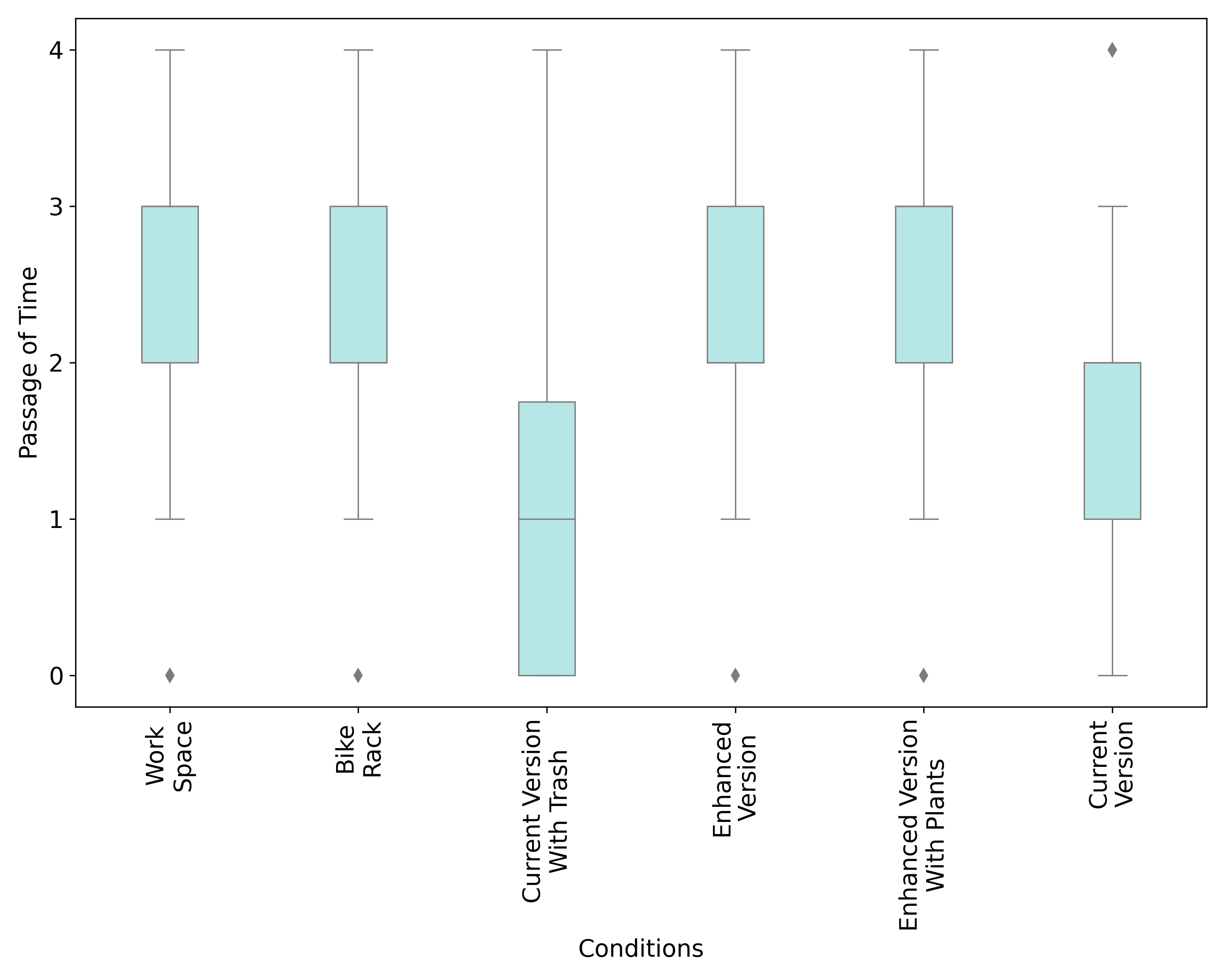}
	\caption{Comparison of perception of the passage of time across different conditions}
	\label{passage of time}
\end{figure}

As shown in Table \ref{time table}, the addition of a workspace to the cabin has the highest $\beta$ estimate ($\beta$=1.285), implying that participants perceive a faster passage of time compared to the current version. Conversely, the presence of trash in the cabin has a significant negative $\beta$ estimate ($\beta$= -0.586), indicating a slower passage of time in comparison to the current version of public transportation. The perceived passage of time by participants was similar in the four conditions of enhanced version, enhanced version with plants, bike racks, and workspace. However, the addition of trash differed from the current version of public transportation (Fig. \ref{passage of time}).

\begin{table}[!h]
\caption {The effect of different designs of public transportation space on participants' perception of the passage of time, evaluated with respect to the current version.}\label{time table} 
\resizebox{0.47\textwidth}{!}{%
\begin{tabular}{llllllll}
Variable                     & Estimate & std.error & t-value & df     & Pr(\textgreater{}|t|) & 2.5 \% CI & 97.5 \% CI \\ \hline
Current Version (Intercept)                  & −0.524   & 0.1155    & -4.5343 & 380.31 & 7.75E-06              & −0.751    & −0.297     \\
Current Version With Trash   & −0.586   & 0.0606    & -9.6656 & 1505   & 1.75E-21              & −0.705    & −0.467     \\
Bike Rack                    & 0.662    & 0.0606    & 10.922  & 1505   & 8.92E-27              & 0.543     & 0.781      \\
Work Space                   & 1.285    & 0.0606    & 21.188  & 1505   & 2.09E-87              & 1.166     & 1.404      \\
Enhanced Version             & 0.924    & 0.0606    & 15.236  & 1505   & 7.45E-49              & 0.805     & 1.043      \\
Enhanced Version With Plants & 1.043    & 0.0606    & 17.201  & 1505   & 1.11E-60              & 0.924     & 1.162      \\
Gender                       & −0.101   & 0.0542    & -1.8694 & 299    & 6.25E-02              & −0.208    & 0.005      \\
Ethnicity                    & 0.152    & 0.059     & 2.5767  & 299    & 1.05E-02              & 0.036     & 0.268     
\end{tabular}%
}
\end{table}

\subsubsection{Price}
\begin{figure}[!h]
	\centering
	\includegraphics[width=0.47\textwidth]{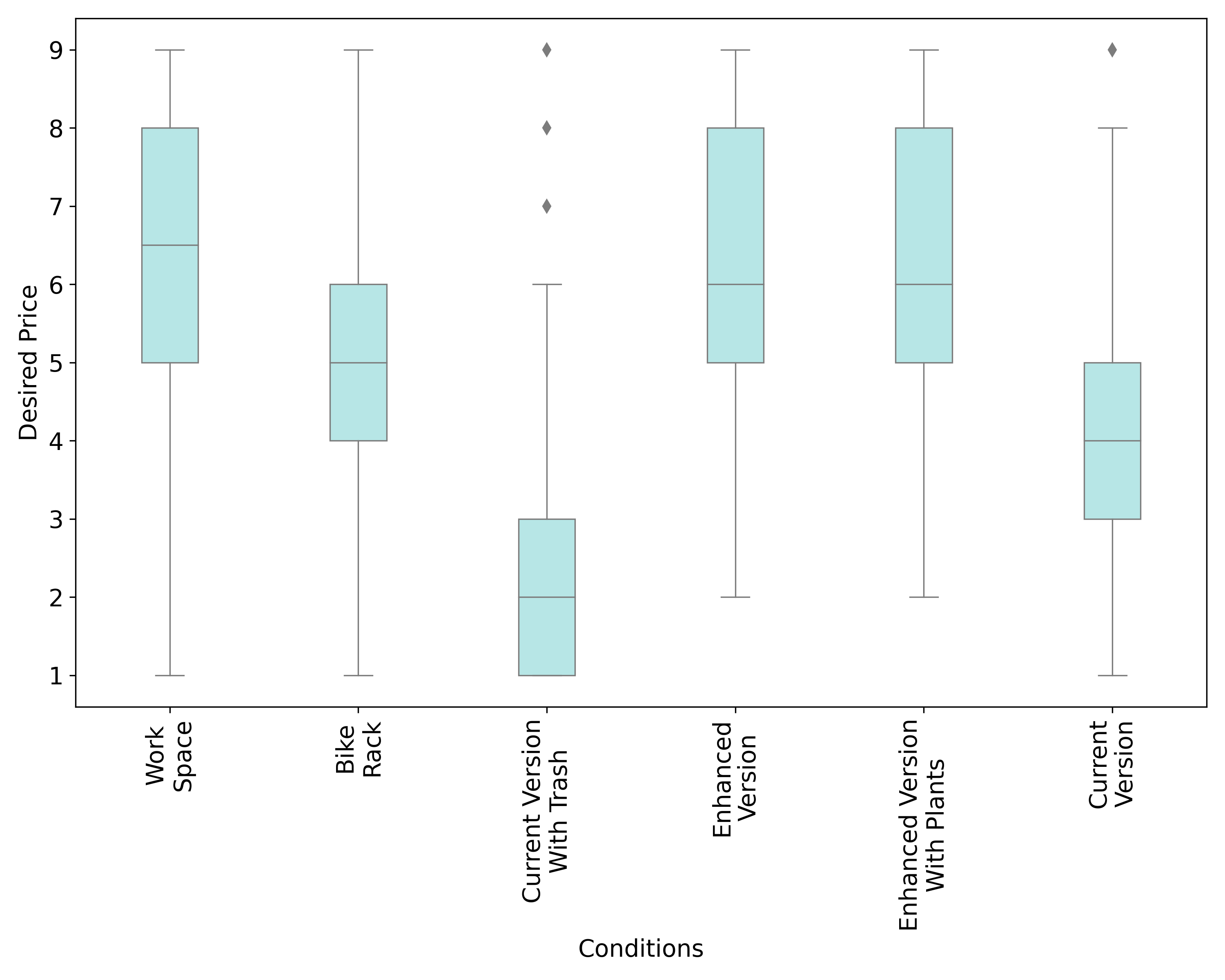}
	\caption{Comparison of reasonable price for every different condition}
	\label{price}
\end{figure}

Participants’ perception of reasonable price significantly varied across different conditions: enhanced version, enhanced version with plants, bike racks, workspace, and current version with added trash, compared to the current version of transportation interior space design (Fig. \ref{price}). As shown in Table \ref{price table}, an enhanced public transportation cabin design with plants increases participants' perception of the reasonable price compared to the current cabin design, with the highest $\beta$ estimate ($\beta$=2.768). Additionally, the addition of workspace ($\beta$=2.606) and having an enhanced version of cabin design ($\beta$=2.493) also have significantly high $\beta$ values, implying a higher price perception among participants compared to the current version of cabin design. Conversely, the presence of trash in the cabin environment has a significant negative $\beta$ estimate ($\beta$=-1.536), indicating a decrease in the perception of reasonable price. Table \ref{price table} summarizes the $\beta$ values and their statistical significance for each condition.  

\begin{table}[!h]
\caption {The effect of different public transportation interior space design conditions on participants’ reasonable price
perception, evaluated with respect to the current version.}\label{price table} 
\resizebox{0.47\textwidth}{!}{%
\begin{tabular}{llllllll}
Variable                     & Estimate & std.error & t-value & df     & Pr(\textgreater{}|t|) & 2.5 \% CI & 97.5 \% CI \\ \hline
Current Version (Intercept)  & −0.810   & 0.2658    & -3.0483 & 350.74 & 2.48E-03              & −1.333    & −0.287     \\
Current Version With Trash   & −1.536   & 0.1145    & -13.416 & 1505   & 7.54E-39              & −1.761    & −1.312     \\
Bike Rack                    & 1.361    & 0.1145    & 11.884  & 1505   & 3.40E-31              & 1.136     & 1.586      \\
Work Space                   & 2.606    & 0.1145    & 22.756  & 1505   & 9.39E-99              & 2.381     & 2.831      \\
Enhanced Version             & 2.493    & 0.1145    & 21.772  & 1505   & 1.37E-91              & 2.269     & 2.718      \\
Enhanced Version With Plants & 2.768    & 0.1145    & 24.172  & 1505   & 2.42E-109             & 2.544     & 2.993      \\
Gender                       & −0.250   & 0.1273    & -1.9633 & 299    & 5.05E-02              & −0.500    & 0.001      \\
Ethnicity                    & −0.028   & 0.1386    & -0.1996 & 299    & 8.42E-01              & −0.301    & 0.245     
\end{tabular}%
}
\end{table}

\subsubsection{Punctuality}
\begin{figure}[!h]
	\centering
	\includegraphics[width=0.47\textwidth]{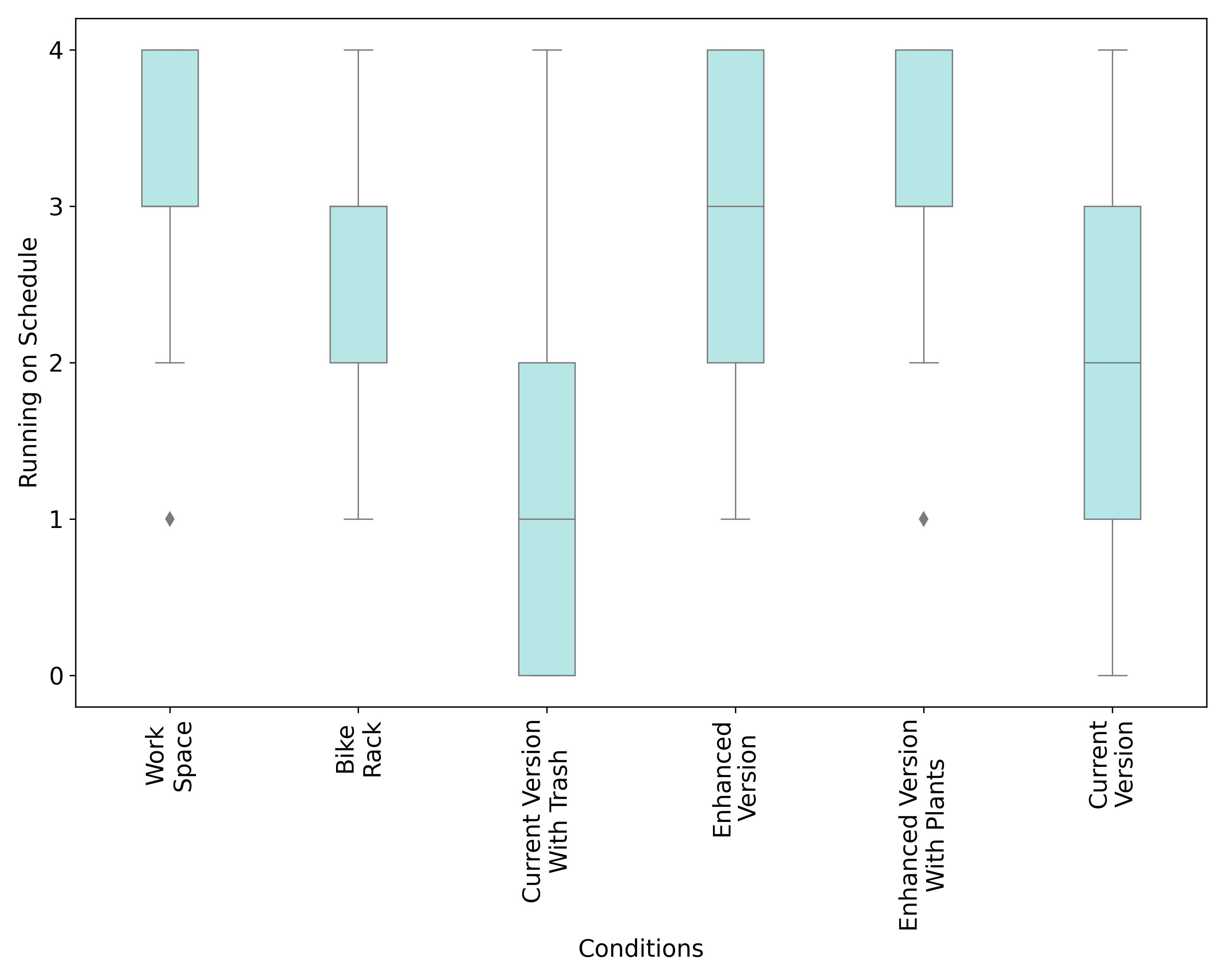}
	\caption{Comparison of perceived punctuality level across different conditions}
	\label{punctuality}
\end{figure}

Similar to price, the perceived punctuality of public transportation containing these cabins was significantly different across the five conditions: enhanced version, enhanced version with plants, bike racks, workspace, and current version with added trash, compared to the current version of public transportation interior space, as depicted in Fig. \ref{punctuality}. Having a workspace in the cabin has the highest estimated $\beta$ ($\beta$=1.215), indicating higher perceived punctuality of the vehicle carrying the cabin compared to the punctuality level of the current public transportation as depicted in Table \ref{punctuality table}. 

\begin{table}[!h]
\caption {The effect of different design of the built environment of public transportation on participants’ punctuality
perception of the train, evaluated with respect to the current version.}\label{punctuality table} 
\resizebox{0.47\textwidth}{!}{%
\begin{tabular}{llllllll}
Variable                     & Estimate & std.error & t-value & df     & Pr(\textgreater{}|t|) & 2.5 \% CI & 97.5 \% CI \\ \hline
Current Version (Intercept)                  & 0.018    & 0.1308    & 0.1358  & 361.75 & 8.92E-01              & −0.239    & 0.275      \\
Current Version With Trash   & −0.805   & 0.0614    & -13.11  & 1505   & 2.91E-37              & −0.925    & −0.684     \\
Bike Rack                    & 0.781    & 0.0614    & 12.733  & 1505   & 2.42E-35              & 0.661     & 0.902      \\
Work Space                   & 1.215    & 0.0614    & 19.8    & 1505   & 9.93E-78              & 1.095     & 1.336      \\
Enhanced Version             & 0.947    & 0.0614    & 15.43   & 1505   & 5.59E-50              & 0.827     & 1.067      \\
Enhanced Version With Plants & 1.079    & 0.0614    & 17.588  & 1505   & 4.03E-63              & 0.959     & 1.2        \\
Gender                       & −0.008   & 0.0621    & -0.1268 & 299    & 8.99E-01              & −0.130    & 0.114      \\
Ethnicity                    & 0.016    & 0.0677    & 0.2324  & 299    & 8.16E-01              & −0.117    & 0.149     
\end{tabular}%
}
\end{table}

\subsubsection{Likelihood to recommend to others}

\begin{figure}[!h]
	\centering
	\includegraphics[width=0.47\textwidth]{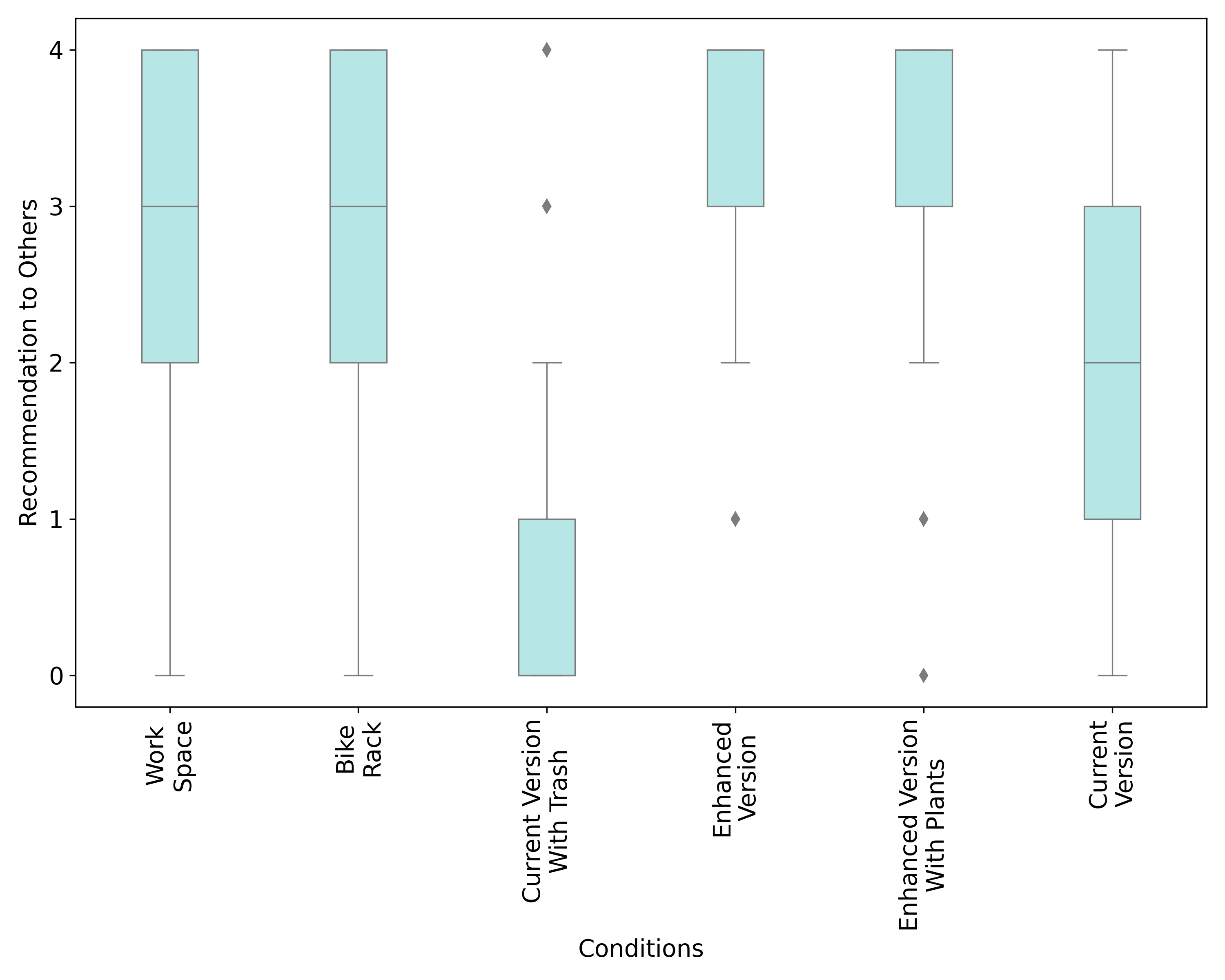}
	\caption{Comparison of the potential to recommend these cabins to others across different conditions}
	\label{recommendation to others}
\end{figure}

The potential for recommending public transportation containing these cabins to others varied significantly across different conditions: enhanced version, enhanced version with plants, bike racks, workspace, and current version with added trash, compared to the current version of public transportation interior space, as depicted in Fig. \ref{recommendation to others}. As shown in Table \ref{recom table}, the enhanced version of cabin design with plants has the highest $\beta$ value ($\beta$=1.636), indicating an increase in participants' likelihood of recommending the cabin to others compared to the current version of public transportation. Conversely, the current version with added trash has a negative $\beta$ value ($\beta$=-1.377), implying a lower potential among participants to recommend the space to others. 

\begin{table}[!h]
\caption {The effect of different transportation interior space designs on participants’ potential to recommend to others, evaluated with respect to the current version.}\label{recom table} 
\resizebox{0.47\textwidth}{!}{%
\begin{tabular}{lrrrrrrr}
Variable                     & \multicolumn{1}{l}{Estimate} & \multicolumn{1}{l}{std.error} & \multicolumn{1}{l}{t-value} & \multicolumn{1}{l}{df} & \multicolumn{1}{l}{Pr(\textgreater{}|t|)} & \multicolumn{1}{l}{2.5 \% CI} & \multicolumn{1}{l}{97.5 \% CI} \\ \hline
Current Version (Intercept)  & -0.064                       & 0.130                         & -0.495                      & 388.32                 & 6.21E-01                                 & -0.320                        & 0.191                          \\
Current Version With Trash   & -1.377                       & 0.071                         & -19.385                     & 1505                   & 6.55E-75                                 & -1.517                        & -1.238                         \\
Bike Rack                    & 0.788                        & 0.071                         & 11.091                      & 1505                   & 1.57E-27                                 & 0.649                         & 0.927                          \\
Work Space                   & 1.182                        & 0.071                         & 16.636                      & 1505                   & 3.54E-57                                 & 1.043                         & 1.322                          \\
Enhanced Version             & 1.513                        & 0.071                         & 21.296                      & 1505                   & 3.55E-88                                 & 1.374                         & 1.653                          \\
Enhanced Version With Plants & 1.636                        & 0.071                         & 23.020                      & 1505                   & 1.04E-100                                & 1.496                         & 1.775                          \\
Gender                       & -0.077                       & 0.061                         & -1.269                      & 299                    & 2.06E-01                                 & -0.196                        & 0.042                          \\
Ethnicity                    & 0.041                        & 0.066                         & 0.619                       & 299                    & 5.36E-01                                 & -0.089                        & 0.171                          \\
\end{tabular}%
}
\end{table}

\subsubsection{Workspace utility}
\begin{figure}[!h]
	\centering
	\includegraphics[width=0.40\textwidth]{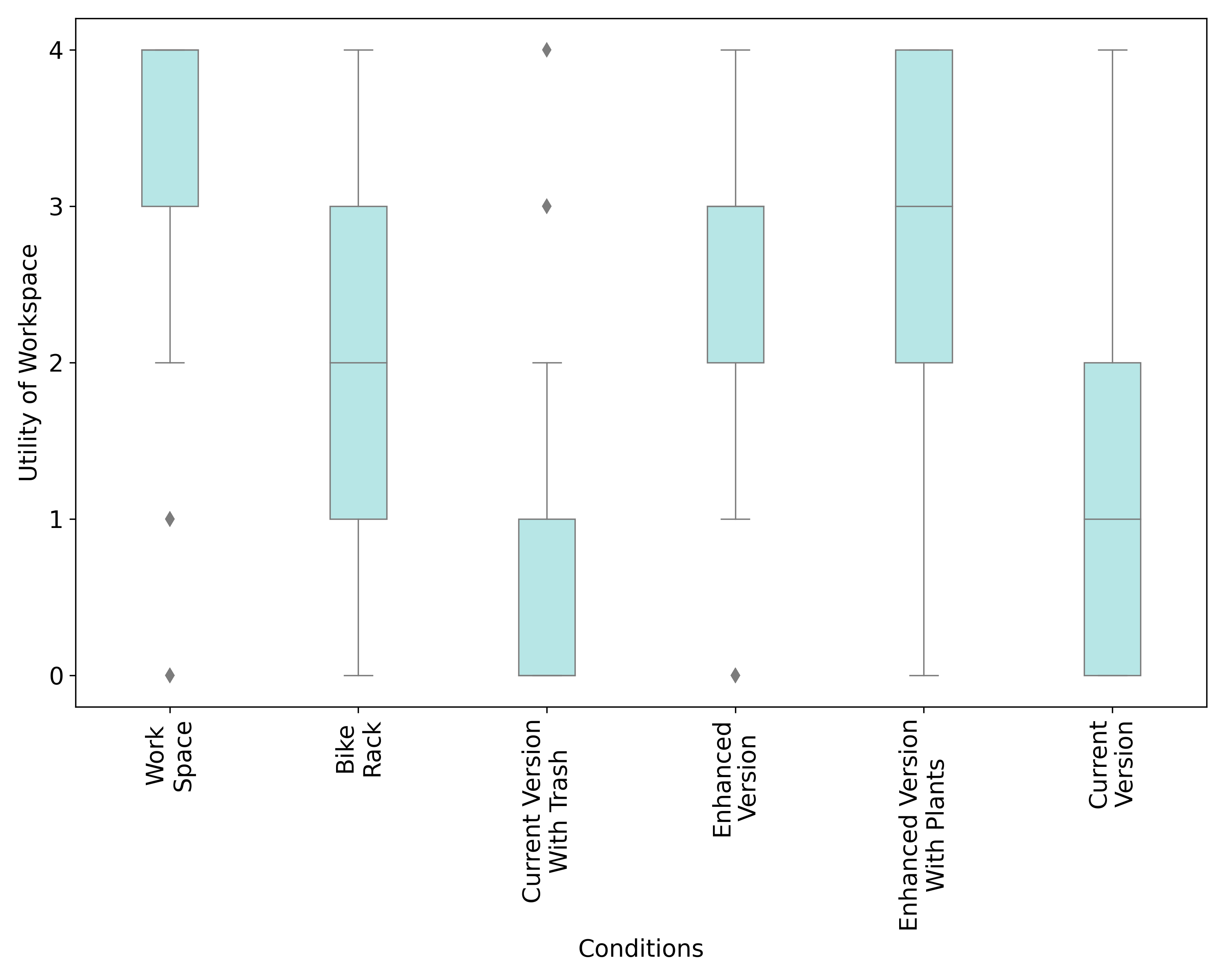}
	\caption{Comparison of tendency to work in the cabin across different conditions}
	\label{utility}
\end{figure}

Participants’ perceptions of workspace utility significantly varied across different conditions, including enhanced version, enhanced version with plants, bike racks, workspace, and current version with added trash, compared to the current version of public transportation interior space design, as depicted in Fig. \ref{utility}. As shown in Table \ref{utility table}, the presence of a workspace in the cabin has the highest estimated $\beta$ ($\beta$=1.907), indicating an increased tendency among participants to work in this cabin while commuting compared to the current version of public transportation cabin design.   

\begin{table}[!h]
\caption {The effect of different public transportation interior space designs on participants’ tendency to work while commuting, evaluated with respect to the current version.}\label{utility table}
\resizebox{0.47\textwidth}{!}{%
\begin{tabular}{llllllll}
Variable                     & Estimate & std.error & t-value & df     & Pr(\textgreater{}|t|) & 2.5 \% CI & 97.5 \% CI \\ \hline
Current Version (Intercept)  & −0.702   & 0.1716    & -4.0922 & 356.57 & 5.29E-05              & −1.040    & −0.365     \\
Current Version With Trash   & −0.745   & 0.0775    & -9.6102 & 1505   & 2.91E-21              & −0.897    & −0.593     \\
Bike Rack                    & 0.599    & 0.0775    & 7.7309  & 1505   & 1.94E-14              & 0.447     & 0.751      \\
Work Space                   & 1.907    & 0.0775    & 24.602  & 1505   & 1.30E-112             & 1.755     & 2.059      \\
Enhanced Version             & 1.391    & 0.0775    & 17.939  & 1505   & 2.29E-65              & 1.239     & 1.543      \\
Enhanced Version With Plants & 1.646    & 0.0775    & 21.228  & 1505   & 1.08E-87              & 1.494     & 1.798      \\
Gender                       & −0.052   & 0.0818    & -0.6398 & 299    & 5.23E-01              & −0.213    & 0.109      \\
Ethnicity                    & −0.021   & 0.0891    & -0.2399 & 299    & 8.11E-01              & −0.197    & 0.154     
\end{tabular}%
}
\end{table}

\subsection{Eye-tracking Metrics}
\begin{figure}[htbp]
    \centering
    \begin{subfigure}[b]{0.3\textwidth}
        \includegraphics[width=\textwidth]{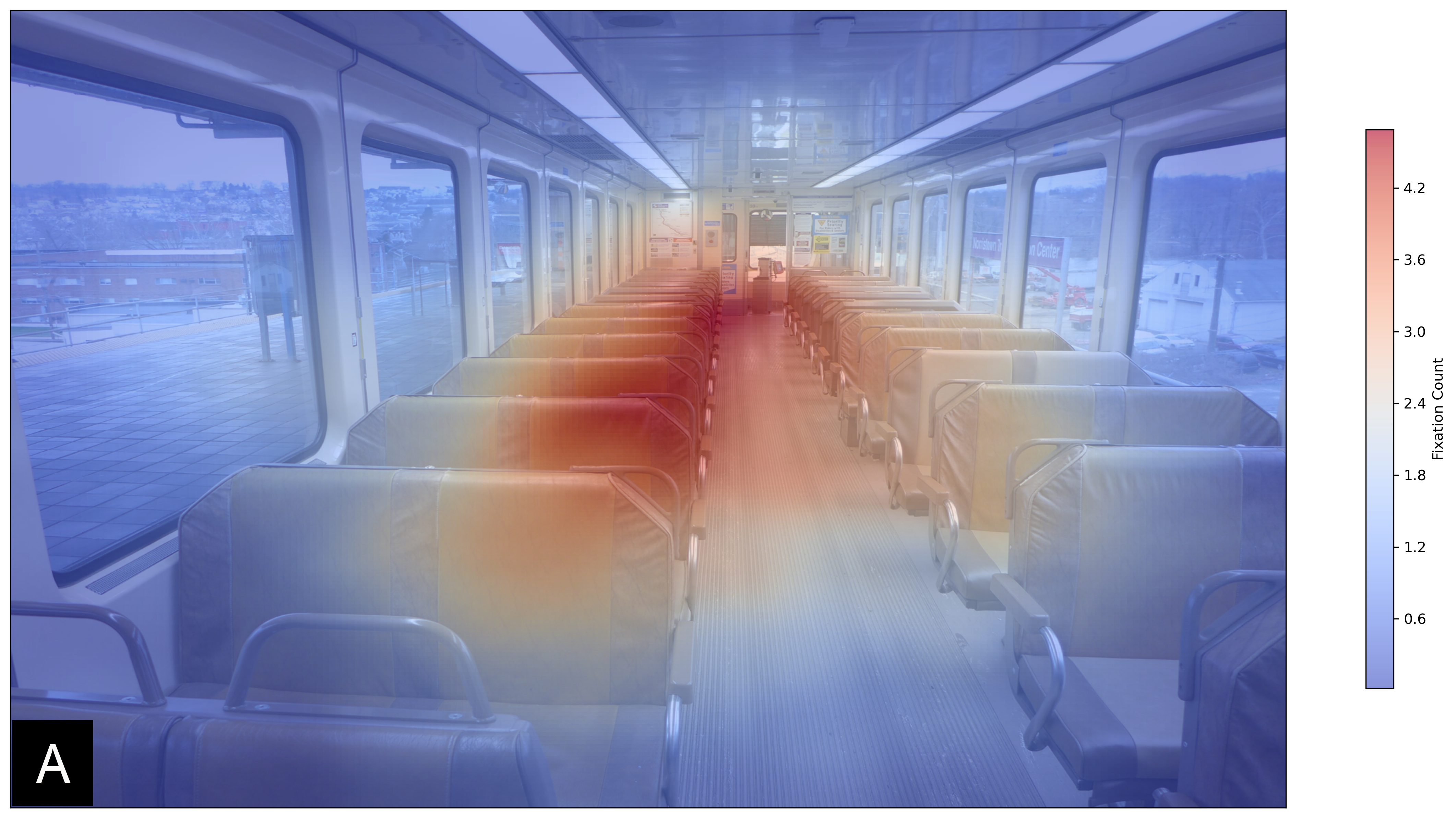}
        \label{baseline}
    \end{subfigure}
    \hfill
    \begin{subfigure}[b]{0.3\textwidth}
        \includegraphics[width=\textwidth]{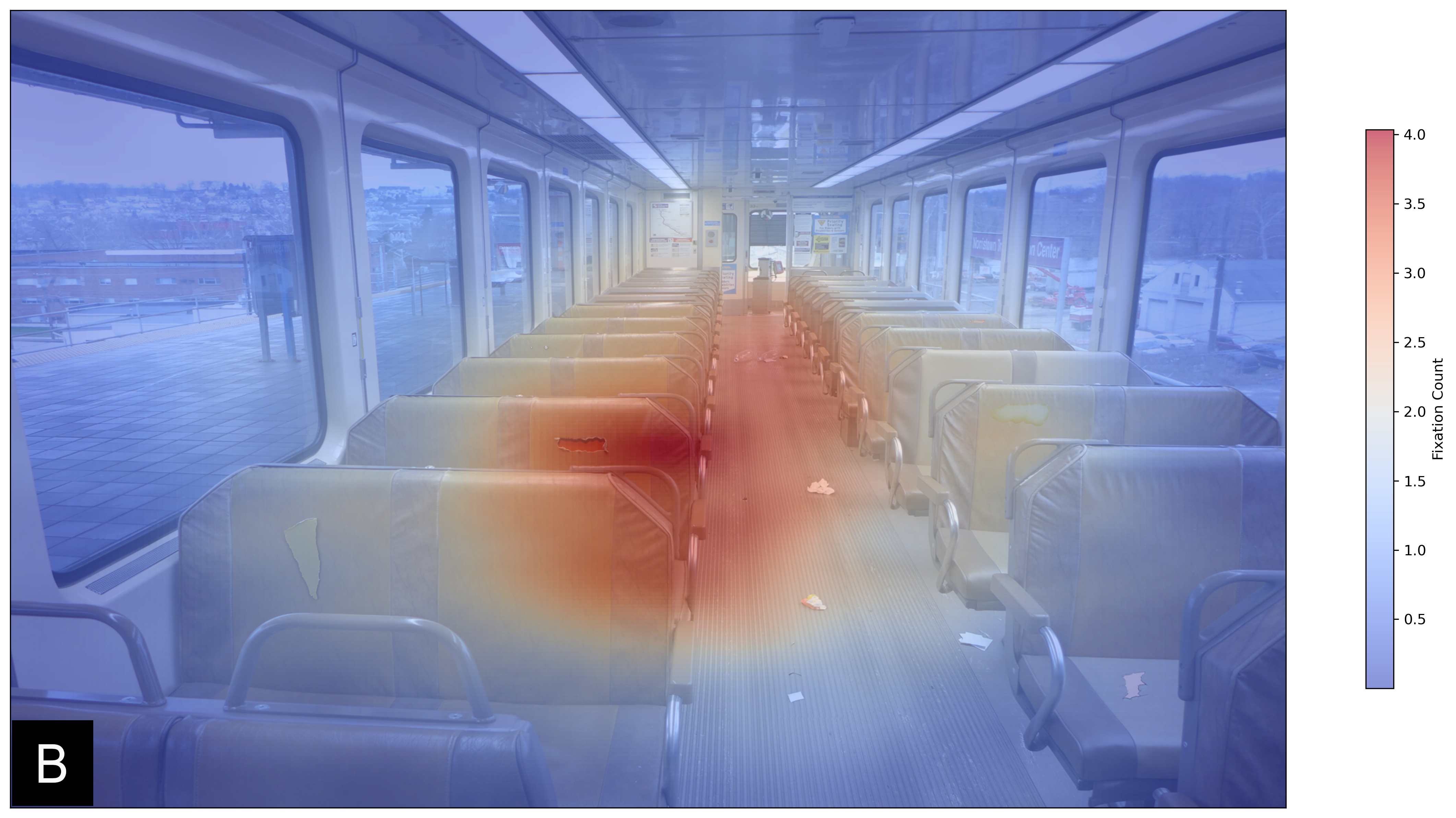}
        \label{trash}
    \end{subfigure}
    \hfill
    \begin{subfigure}[b]{0.3\textwidth}
        \includegraphics[width=\textwidth]{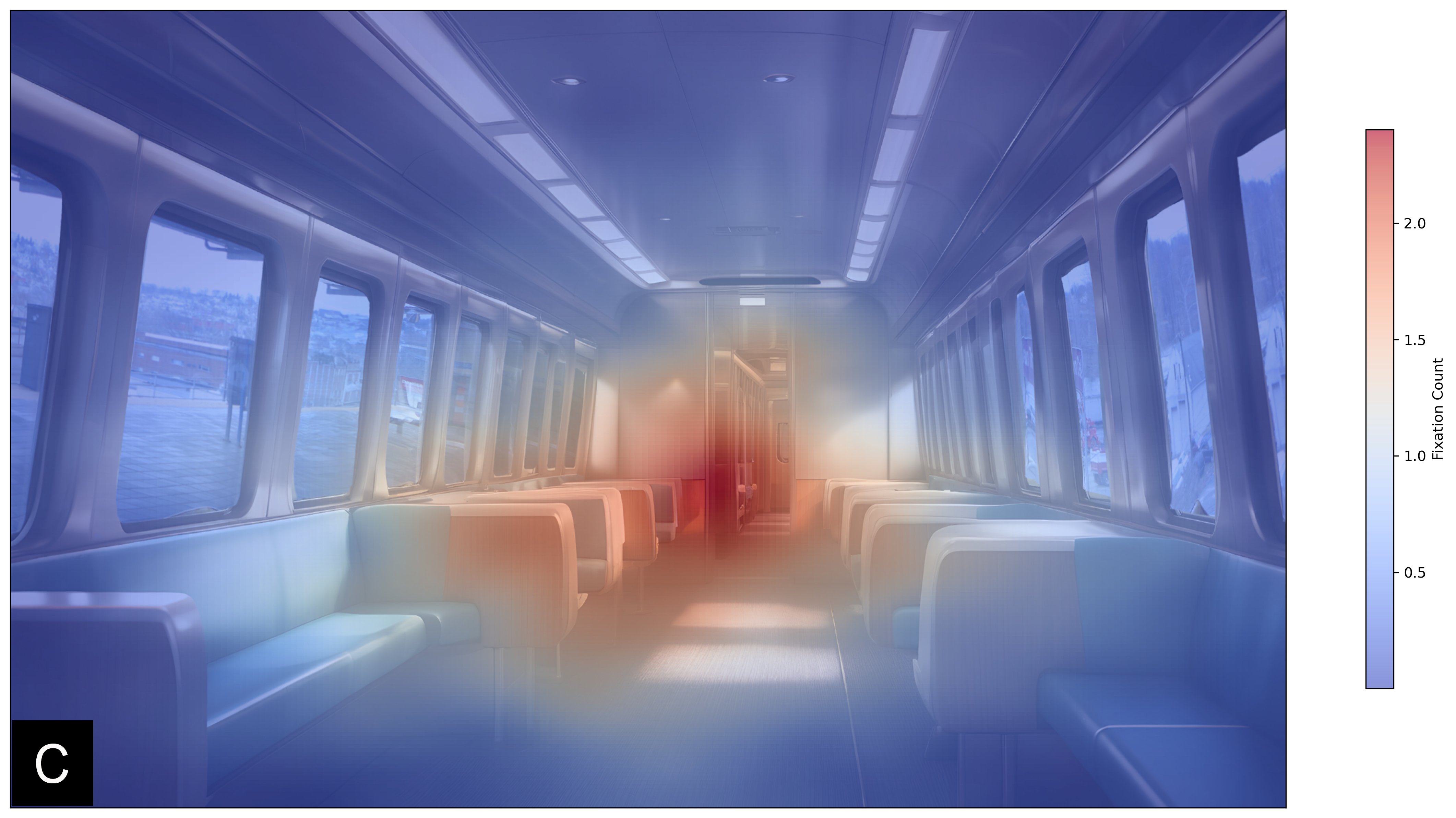}
        \label{enhanced}
    \end{subfigure}
    \begin{subfigure}[b]{0.3\textwidth}
        \includegraphics[width=\textwidth]{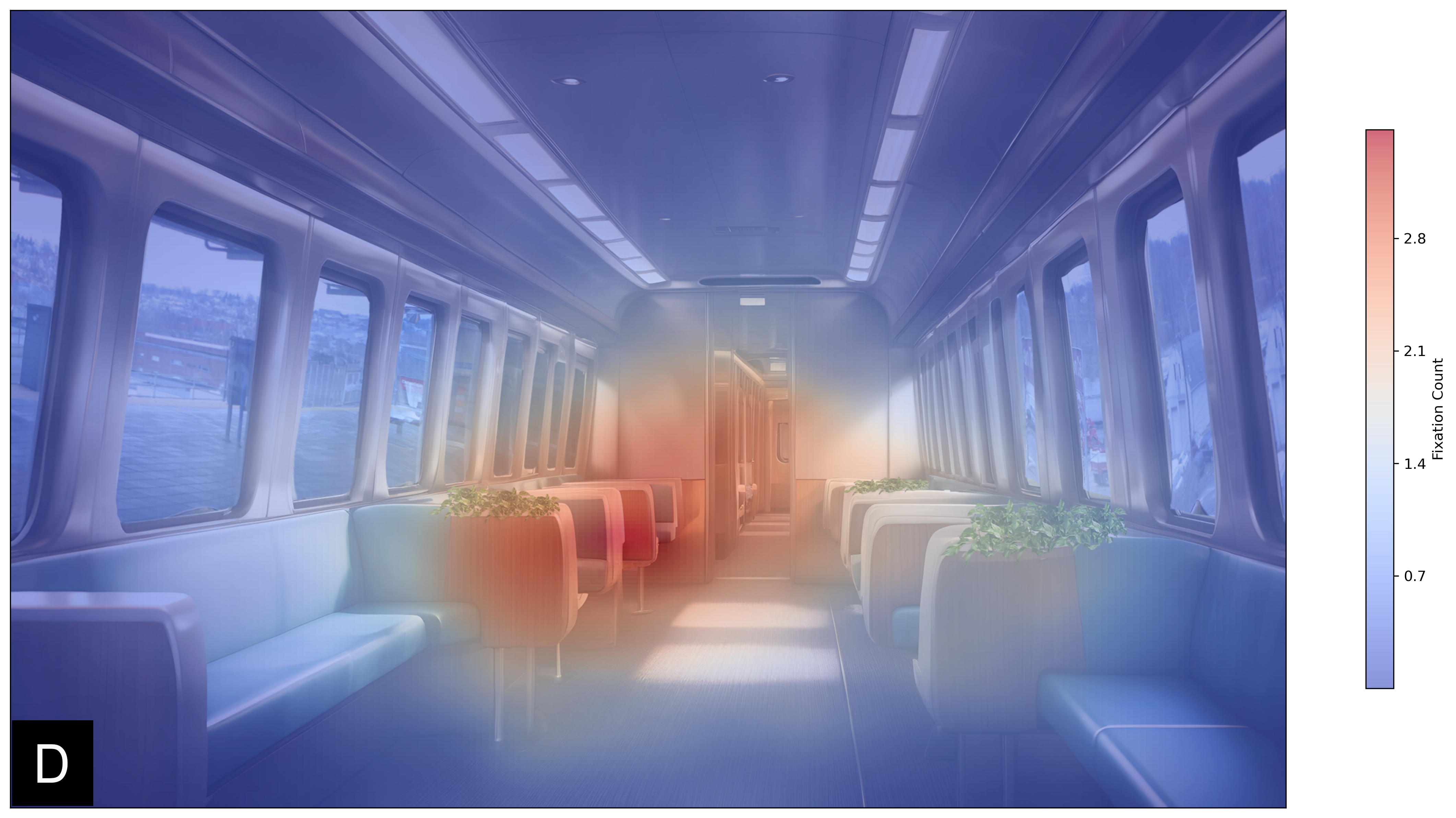}
        \label{plants}
    \end{subfigure}
    \hfill
    \begin{subfigure}[b]{0.3\textwidth}
        \includegraphics[width=\textwidth]{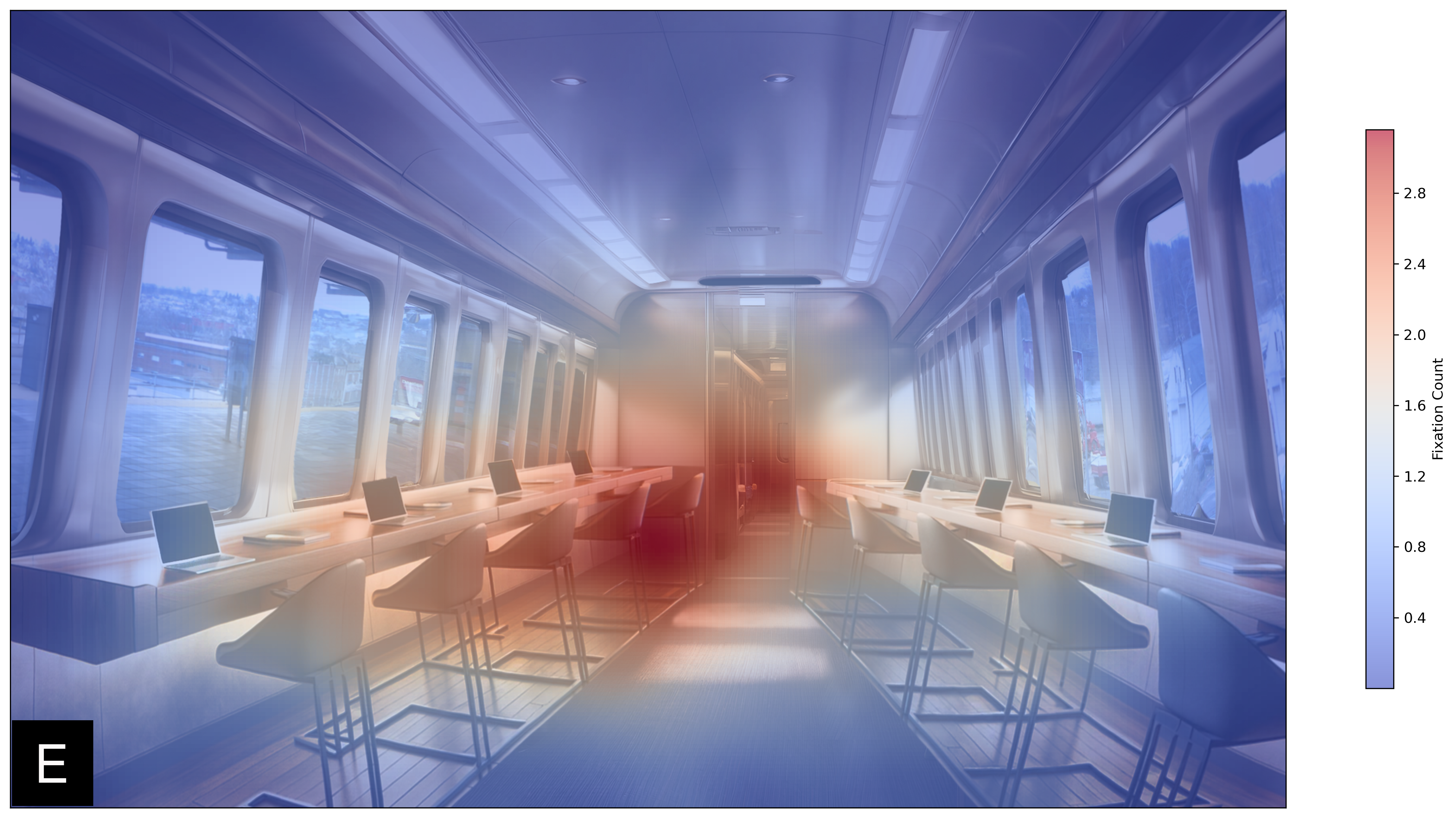}
        \label{work}
    \end{subfigure}
    \hfill
    \begin{subfigure}[b]{0.3\textwidth}
        \includegraphics[width=\textwidth]{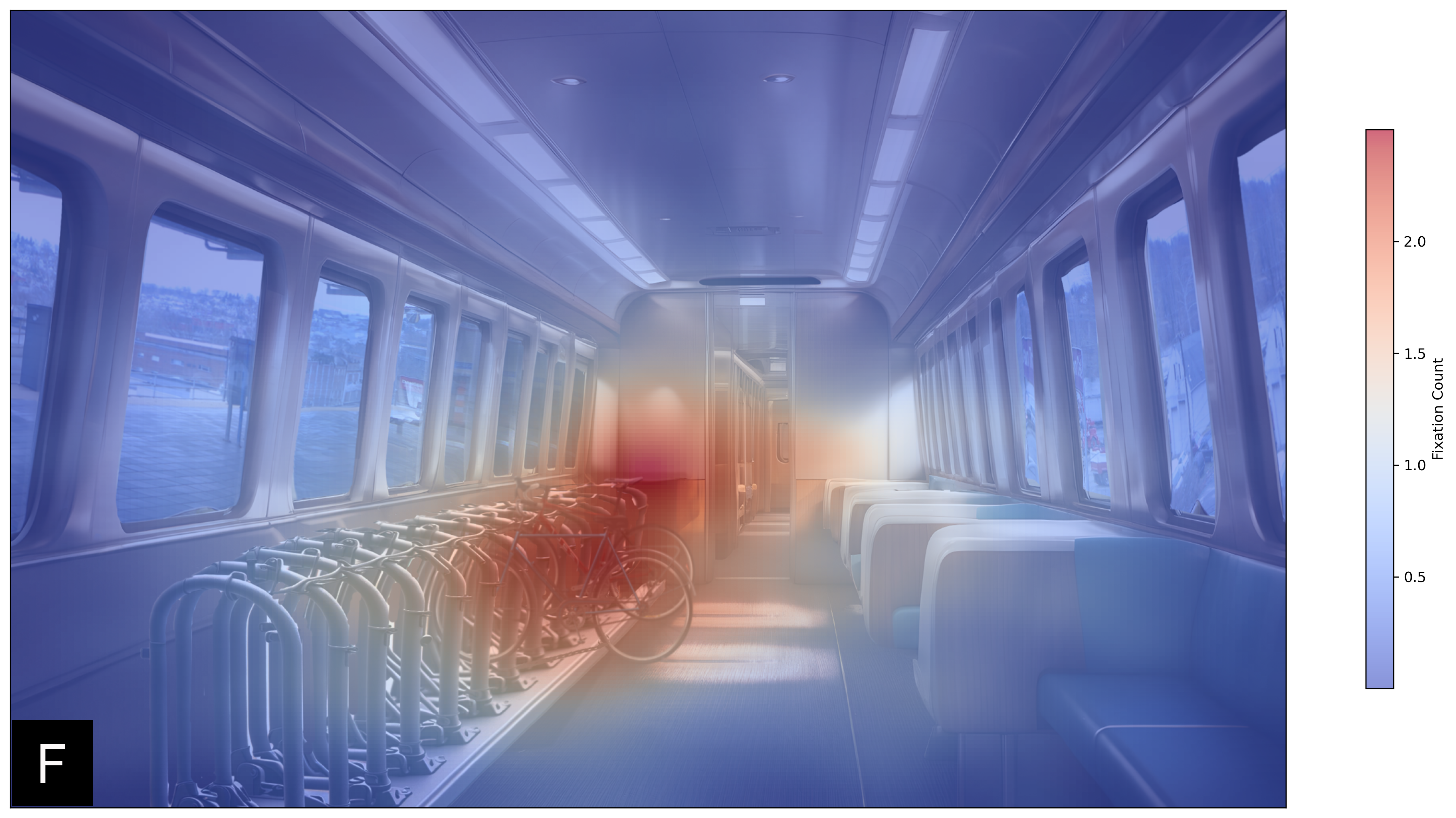}
        \label{bike}
    \end{subfigure}
    
    \caption{The heatmap shows the number of fixations in different train cabins: (A) Current version, (B) Current version with visible trash added, (C) Enhanced version, (D) Enhanced version with plants, (E) Cabin containing workspace, and (F) Bike rack.}
    \label{fig:mainfig}
\end{figure}

For brevity, while we do not focus on the details of eye-tracking metrics in this article, a preliminary analysis of the heatmaps of participants' eye gaze fixations shows that our designs were consciously perceived by the participants. More specifically, in Fig. \ref{fig:mainfig}, we observe that the participants' attention was focused on the bikes, workspace equipment, and plants themselves, even though in all of the pictures, people tended to look toward the middle left of the image.

\section{Discussion}

Our study focused on quantifying the effect of the alternative designs of a transportation cabin on participants' well-being, and various perceptual metrics such as perceptions of safety, reasonable cost associated with it, punctuality, perception of comfort, utilization of workspace, and likelihood of recommending it to others. Considering many people spend a significant portion of their time within these cabins, understanding the effect of these factors can be very helpful in enhancing community well-being. The alternative designs included the (1) current version, (2) current version with added trash and low upkeep and maintenance, (3) enhanced version from a material and lighting point of view, (4) enhanced version with plants and biophilic elements, (5) with added workspace, and (6) with a bike rack. 

\subsection{Well-being metrics}

Our study reveals that cabin maintenance and upkeep significantly influence passenger well-being from various metrics of interest and can result in decrementing public well-being, especially those who use such spaces regularly. More specifically, we observed that the presence of trash or unmaintained wear and tear on the seats in current cabins notably diminishes well-being compared to those without trash, resulting in increased stress levels and more negative emotional valence. These findings underscore the importance of maintaining clean public transportation spaces to enhance community well-being, particularly for regular users. Our research is consistent with prior studies in urban planning showing that poorly maintained urban environments, including public transportation systems, can have a detrimental effect on mental health and overall quality of life \cite{UrbanBlight2017,Schultz2013}, emphasizing the significance of cleanliness in public transportation systems for promoting passenger satisfaction and health \cite{Li2020,Braubach2017}.

All of the new designs (i.e., enhanced version, enhanced version with plants, bike rack, and workspace) had a significant positive effect on participants' stress levels, reducing overall stress compared to the current version. This observation indicates that aesthetically enhancing and maintaining the cabin can contribute to reducing stress among commuters compared to the current version of the public transportation spaces, which is similar to findings from Ka Ho Tsoi et al. survey, showing that poor station environments and uncomfortable train compartments are among the primary stressors for daily commuters \cite{TSOI2023103833}. More broadly, similar studies in the urban development research arena also show that the space characteristics such as dislike for the appearance of one's estate or road, indicative of negative perceptions of design and maintenance, had a negative impact on mental health \cite{GUITE20061117}. 


Interestingly, arousal levels remained relatively constant across all conditions, indicating that stress and negative emotions fluctuated, while general alertness did not. This consistency in arousal suggests that passengers' baseline level of engagement or alertness was maintained regardless of the environmental conditions. This can indicate that arousal might be influenced more by individual traits or external factors not addressed by cabin environment modifications. Additionally, participants may have had difficulty interpreting the question related to arousal due to its more abstract nature (relative to valence), especially in the context of cabin space. 

The lack of maintenance and presence of trash significantly diminished participants' creativity, highlighting the constraining effect of lack of maintenance on self-reported cognitive processes. While all the designs (bike rack, enhanced version, enhanced version with plants, and workspace) enhanced participants' perception of creativity, their ratings were lower than the pre-experiment assessment, except for the workspace design, which matched the pre-experiment level. These findings are in line with previous research showing the adverse impact of clutter on people's well-being \cite{Crum2019a,Crum2019b}. Notably, the only design that aligned with participants' prior creativity level was the design with a designated workspace, which is similar to the findings in research in the relationship between adult creativity and the environment \cite{LEE2023101276}, indicating the importance of such elements for enhanced creativity among commuters, especially in the long-distance trains.

Also, the results showed that gender had a significant effect 

Incorporating biophilic elements into the cabin design yielded the most substantial improvements in stress and positive emotions when set against the current version of cabins. Stress levels were the lowest among all conditions, and participants experienced significantly more positively valenced emotions. Creativity was also notably enhanced, reflecting the cognitive benefits of natural elements. This enhancement has been higher than other improved conditions in the experiment. These findings support prior research on the advantages of biophilic design and extend its application to public transportation, showing that integrating natural elements like plants can markedly improve well-being \cite{ZHONG2022114,Jha2022}.


\subsection{Transportation space metrics}

In the cabin with lower maintenance and the presence of visible trash, participants expressed lower levels of perceived cleanliness, comfort, and safety. They also perceived time as passing more slowly, felt that the environment offered less value for money, and rated the service as less punctual. This particular cabin was the least likely to be recommended to others, and its workspace utility was minimal. These results reinforce previous findings regarding users' significant unfavorable perceptions of spaces with low upkeep \cite{Mouratidis2021} and show that such environments can result in decreasing public transportation usage.

In the enhanced version cabin, perceptions of cleanliness and comfort were higher, and participants felt safer. The passage of time was perceived as faster, and the cabin was seen as providing better value for money. Punctuality ratings were higher, reflecting increased satisfaction with the service. This version was more likely to be recommended to others, and the workspace utility was positively rated, suggesting that not only can design enhance the overall passenger experience, but they can result in enhancing public transportation usage, and the domino effect of recommending it to others. These results suggest that public transportation authorities can enhance the public transit share of the transportation by maintaining the spaces.

Also, in the enhanced version with biophilic elements cabin, perceptions of cleanliness, comfort, and safety were the highest, and participants felt the safest in this environment compared to the current version. The passage of time was perceived as faster, and the cabin was viewed as offering excellent value for money. Punctuality ratings were higher, and this version was the most likely to be recommended to others. Workspace utility was also highly rated, demonstrating the overall benefits of biophilic design. Such results reverberate previous findings on the positive impact of plants in the scene on the subjective perception of space \cite{doi:10.1080/17508975.2020.1732859}. Additionally, these results are in line with the studies from other built environments such as offices where biophilic elements were shown to enhance occupants' perception of the environment and well-being metrics \cite{Larsen1998}.

The addition of a bike rack to the cabin also positively influenced passengers' perception of public transportation-related metrics. Participants reported higher perceptions of cleanliness, comfort, and safety ratings, as well as a faster passage of time. The bike rack cabin was perceived as offering good value for money, and punctuality ratings were higher. This version was more likely to be recommended to others, and the workspace utility was positively rated, indicating that practical amenities can enhance passenger satisfaction.

In the workspace cabin, perceptions of cleanliness and comfort improved. Safety ratings were higher, and participants reported a faster passage of time. The workspace cabin was viewed as offering excellent value for money, and punctuality ratings were higher compared to the current version. This version was highly recommended to others, and the workspace utility was the highest among all conditions, highlighting the benefits of providing functional spaces for passengers who may work during travel.

The preliminary results of the eye-tracking and heatmaps indicate that the new designs captured the majority of participants' attention. Notably, in the bike rack cabin, the bikes themselves were the primary focus. Additionally, workspace and plants also drew significant attention. While participants tended to look towards middle to the left of the images initially, which is referred to as ``center fixation bias" \cite{Tatler2007}, the new designs, including the bike rack, workspace, seat arrangement, biophilic design elements, low-maintenance features, and visible wear and tear, attracted more attention in the areas with those elements compared to the current version, qualitatively.

\section{Limitation}
This study has several limitations that may impact the generalizability of this work as follows: The participants in the study were all from the Northeastern region of the United States. In order to improve the accuracy and applicability of the results, future research should include a wider geographical area. The online format of the study meant that we could not control the participants' environmental conditions, which could have affected their responses. For example, people who are taking the survey in stressful environments might be affected by their environment of study and may report higher stress levels, highlighting a limitation of online surveys. We have partly addressed this issue by adding baseline questions. 

Online surveys can introduce bias through participant selection. This is due to the use of an online platform to recruit participants, leading to a sample population consisting only of individuals who are comfortable being surveyed on websites and having their eye-gaze captured via their webcams. This could negatively impact the generalizability of our conclusions. Therefore, future efforts should focus on recruiting individuals from diverse recruitment platforms to capture a broader range of data. Additionally, having a lengthy survey may lead to participant exhaustion, causing them to answer the latter part of the questionnaire reluctantly. It should be noted that we have used counterbalancing in our design to distribute potential fatigue effects equally among conditions, mitigating the risk that participant exhaustion will bias the results. While we piloted the study to ensure the time taken to finish it is commensurate with the payment and is not exhausting, this issue remains as one of the important factors affecting survey studies.  

Lastly, while we did not focus specifically on eye-tracking data in this manuscript, it is worth noting that the use of eye-tracking data captured through online platforms has been well-documented in the literature. These studies demonstrate the feasibility and potential of remote eye-tracking for various research purposes. However, a notable limitation is the inability to ensure that respondents' lighting conditions are appropriate for accurate eye-tracking measurements. Inconsistent or inadequate lighting can significantly affect the quality of the eye-tracking data, potentially leading to unreliable results. Future research should address this limitation by implementing guidelines for optimal lighting conditions or developing more robust eye-tracking technologies that can compensate for varying lighting environments.

\section{Conclusion and Future Work}
This study quantified the effect of public transportation spaces on user's well-being metrics and perpetual characteristics. Our study demonstrates that environmental enhancements in public transportation cabins significantly impact passengers' well-being. Clean and well-designed spaces, particularly those incorporating natural elements like plants, not only reduce stress but also improve positive valence, enhance creativity, and increase perceptions of cleanliness, comfort, and safety. These enhancements also lead to a faster-perceived passage of time, better value for money, higher punctuality ratings, a greater likelihood of being recommended to others, and higher workspace utility. Conversely, poorly maintained environments can negatively affect various aspects of well-being, underscoring the critical importance of maintaining cleanliness. These insights suggest that transportation providers should prioritize clean, functional, and aesthetically pleasing designs to enhance passenger satisfaction and promote overall well-being.

Future work should focus on enhancing the generalizability of surveys by employing methods that are more immersive such as virtual environments, and are closer to real-world conditions. Additionally, future work can extend to field studies that expand on this topic within real-world cabin environments. While this study focused on self-reported metrics and objective questionnaires, future work can take advantage of other sensing mechanisms to acquire a deeper understanding of user behavior and state within such environments such as the use of physiological metrics. Part of the future work of this project involves a detailed analysis of the eye-tracking data to gain a deeper understanding of unconscious perceptions and patterns of users while imagining themselves in the designed cabins. Lastly, while this paper has primarily focused on the user aspects, it is important to consider the perspectives of other stakeholders in the transportation industry, such as transit authorities and related entities. Future research should aim to develop user-centered designs by involving multiple parties simultaneously to create more comprehensive solutions.

\bibliographystyle{IEEEtran}
\bibliography{mybibfile}
%



\end{document}